\documentstyle[psfig,aps]{revtex}
\def\be{\begin{equation}}
\def\ee{\end{equation}}
\def\bea{\begin{eqnarray}}
\def\eea{\end{eqnarray}}

\def\IR{{\hbox{{\rm I}\kern-.2em\hbox{\rm R}}}}
\def\IB{{\hbox{{\rm I}\kern-.2em\hbox{\rm B}}}}
\def\IN{{\hbox{{\rm I}\kern-.2em\hbox{\rm N}}}}
\def\IC{\,\,{\hbox{{\rm I}\kern-.59em\hbox{\bf C}}}}
\def\IZ{{\hbox{{\rm Z}\kern-.4em\hbox{\rm Z}}}}
\def\IP{{\hbox{{\rm I}\kern-.2em\hbox{\rm P}}}}
\def\IH{{\hbox{{\rm I}\kern-.4em\hbox{\rm H}}}}
\def\ID{{\hbox{{\rm I}\kern-.2em\hbox{\rm D}}}}
\def\II{{\hbox{\rm I}\kern-.2em\hbox{\rm I}}}

\begin{document}
\title{\begin{flushright}
{\small hep-th/9904197}
\end{flushright}Holography, Thermodynamics and Fluctuations of
Charged AdS Black Holes} \author{Andrew Chamblin$^a$, Roberto
Emparan$^b$, Clifford V. Johnson$^c$ and Robert C. Myers$^d$}
\author{} \address{$^a$ D.A.M.T.P., Silver Street, Cambridge, CB3 9EW,
UK.}  \address{$^b$ Department of Mathematical Sciences, University of
Durham, DH1 3LE, UK.\\ Departamento de F{\'\i}sica Te\'orica,
Universidad del Pa{\'\i}s Vasco, Apdo. 644, E-48080 Bilbao, Spain.}
\address{$^c$ Department of Physics and Astronomy, University of
Kentucky, Lexington, KY 40506--0055, USA.}  \address{$^d$Physics
Department, McGill University, Montr\'{e}al, PQ, H3A 2T8, Canada.}
\address{\small $^a$H.A.Chamblin@dampt.cam.ac.uk,
$^b$Roberto.emparan@durham.ac.uk, $^c$cvj@pa.uky.edu,
$^d$rcm@hep.physics.mcgill.ca}

\author{}

\maketitle \begin{abstract} The physical properties of
Reissner--Nordstr\"om black holes in $(n{+}1)$--dimensional anti--de
Sitter spacetime are related, by a holographic map, to the physics of
a class of $n$--dimensional field theories coupled to a background
global current.  Motivated by that fact, and the recent observations
of the striking similarity between the thermodynamic phase structure
of these black holes (in the canonical ensemble) and that of the van
der Waals--Maxwell liquid--gas system, we explore the physics in more
detail.  We study fluctuations and stability within the equilibrium
thermodynamics, examining the specific heats and electrical
permittivity of the holes, and consider the analogue of the
Clayperon equation at the phase boundaries. Consequently, we
refine the phase diagrams in the canonical and grand canonical
ensembles. We study the interesting physics in the neighbourhood of
the critical point in the canonical ensemble. There is a second order
phase transition found there, and that region is characterized by a
Landau--Ginzburg model with $A_3$ potential. The holographically dual
field theories provide the description of the microscopic degrees of
freedom which underlie all of the thermodynamics, as can be seen by
examining the form of the microscopic fluctuations.
\end{abstract}

\newcommand{\beq}{\begin{equation}}
\newcommand{\eeq}{\end{equation}}
\newcommand{\beqa}{\begin{eqnarray}}
\newcommand{\eeqa}{\end{eqnarray}}
\newcommand{\rp}{r_+}

\section{Introduction and Summary}\label{sec:intro}
Explaining the thermodynamic nature of black holes was recognized
as an essential
hallmark of any complete quantum theory of gravity long before such a
theory was constructed. The semi--classical approach to quantum
gravity, which has become quite a mature subject over the
years\cite{gibbhawk,refined}, allows for the computation of a number
of physical quantities. These treatments ignore the details of how a
specific solution of Einstein's equations (regarded as the effective
low energy truncation of the complete quantum gravity) arises, and
instead perform a quantum treatment of field degrees of freedom
in a fixed classical space--time background.

In that way it was learned that the entropy of Bekenstein\cite{beken}
and the temperature of Hawking\cite{hawk}, for example, fit into an
elegant thermodynamic framework, with questions (such as scattering,
unitarity, {\it etc.},) concerning the underlying microscopic
description ---which we might use to construct the underlying
``Statistical Mechanics''--- best left for the future development of a
quantum theory of gravity.

That future is now here. String theory (and/or ``M--theory'') supplies
a microscopic description of the underlying degrees of freedom upon
which a statistical description of the laws of black hole
thermodynamics can be based. This is true even though we do not yet
have a satisfactory way of writing the theory in all regimes: the
``D--brane calculus''~\cite{dnotes} provides a robust
framework within which to describe many properties of black
holes\cite{amanda}, while in turn being firmly rooted in the dynamical
framework of string duality and, ultimately,
M--theory\cite{duality}. Typically, the description of black holes
(and other important geometrical backgrounds) proceeds by translating
their properties into properties of an auxiliary field theory,
identified as residing on the world--volume of some collection of (D--
or M--) branes.

One of the succinct ways of organizing this microscopic description of
the properties of black holes is via ``AdS
Holography''\cite{juan,gubkleb,edads,edadsii,hologram}. Then, the
thermodynamic properties of black holes in anti--de Sitter space--time
are dual to those of a field theory in one dimension
fewer\cite{edads,edadsii}. The fact that the thermodynamic properties
of the AdS black holes\cite{hawkpage} are organized by an effective
field theory is not implausible, in the light of the fact that AdS
acts like a natural ``box'' (with reflecting walls) which neutralizes
the tendency of gravitational interactions to render a canonical
thermodynamic ensemble unstable. The fact that the effective field
theory is one which does not contain gravity, and that it is actually
a ``holographically'' dual four dimensional gauge theory (with
suitable generalizations beyond $D{=}4$) is another striking example
of the fundamental role that gauge theory plays in duality in various
situations\footnote{See also refs.\cite{methroat,oferthroat,etudes}
for discussion of how this extends to relating the physics of linear
dilaton backgrounds to theories on the world--volumes of NS--branes.}.

That the AdS arena stabilizes the thermodynamics of black holes is
especially apparent when one discovers phase structures completely
analogous to familiar thermodynamic systems from elsewhere in
Nature. Such an example can be found in the
Reissner--Nordstrom--anti--de--Sitter (RNadS) systems in various
dimensions\cite{cejmii}. There, the $(Q,T)$ diagram showing the
thermally stable phases for a fixed charge (canonical) ensemble turns
out to be completely analogous to that of the $(T,P)$ phase diagram of
the liquid--gas system. The structure of the first order phase
transitions, {\it etc.}, are controlled by a ``cusp''
catastrophe\cite{catastrophe}, common in the theory of discontinuous
transitions in thermodynamics and many other fields\footnote{Recently,
the cusp catastrophe has appeared again the AdS/CFT context, in
ref.\cite{klemm}.}. Meanwhile the free energy as a function of
temperature $F(T)$, displays the characteristic ``swallowtail'' shape.

In this paper, we report the results of our further examination of
these structures, exploring the equilibrium thermodynamics more
closely, including the effects of considering electrical stability,
and thermal fluctuations. The similarities noted between the RNadS
physics and that of well--defined systems such as the liquid--gas are
more than mere analogies: {\it We find that everything has a very natural
place in classic equilibrium thermodynamics, as is consistent with a
holographic duality to thermal field theory without
gravity.} Accordingly, using the techniques of equilibrium thermodynamics,
we refine the phase diagrams which we found in ref.\cite{cejmii}
somewhat, and identify the generic physical properties which give rise
to the cusp and swallowtail structures.

As discussed in our previous paper~\cite{cejmii}, the thermodynamics of the 
Reissner--Nordstr\"om black holes in the presence of negative
cosmological constant in various dimensions are pertinent (because of
the holographic map) to the thermodynamics of families of field
theories found on the common world--volume of collections of large
numbers of branes (for example M2-- and  D3--branes), in the situation
where a global background current (or its canonical conjugate charge)
has been switched on and held fixed\footnote{See\cite{cveticetal} for
additional work on how to relate charged AdS black holes to 
string/M-theory.}.

Geometrically this is performed by simply setting the M2-- and
D3--branes rotating equally in each of the available transverse
orthogonal two--planes. The higher dimensional angular momentum
becomes the Maxwell $U(1)$ charge after the Kaluza--Klein reduction on
the (now twisted) sphere, which yields the gauged supergravity.
Obtaining a pure Maxwell term in this way is not possible starting
with the M5--brane, and so the seven dimensional
Einstein--Maxwell--anti--de--Sitter (EMadS$_7$) theory defines at best
a close cousin to the field theory found on the M5--branes'
world--volumes. The dual theory relevant to EMadS$_6$ should be
considered in a similar manner.

A truly rich phase structure for the field theories (with transition
temperatures away from $T{=}0$) is obtained only for finite volume,
which is the case we concentrate on here. Our studies correspond to the
study of black holes with spherical horizons, $S^{n-1}$. The field
theory resides on $\IR{\times}S^{n-1}$. The case of infinite volume
corresponds to black holes with horizons $\IR^{n-1}$, and to field
theory on $\IR^{n}$. This is of course the case which comes from
taking directly the near horizon limit of explicit brane solutions.

As shown in ref.\cite{cejmii}, the results for infinite volume may be
easily obtained as a scaling limit of the results of finite volume,
and so we will not discuss them here. Of course, even though we are in
finite volume for much of our discussion, the thermodynamic limit is
still valid here, because the dual field theory is at large $N$, and a
positive power of $N$ measures the number of degrees of freedom in the
field theory (for example, $N^2$ in the case of gauge theory, for
$n{=}4$ here).

The structure of the paper is as follows: In
section~\ref{sec:theholes} we recall the charged black solutions of
the Einstein--Maxwell--anti--de--Sitter system. We also recall the
results of performing the Euclidean section and ensuring its
regularity. In section~\ref{sec:state}, we translate these results
into a statement of about the relation between the thermodynamic
variables of the black hole system in thermodynamic equilibrium {\it
i.e.,} the ``equation of state''. In section~\ref{sec:ensemble}, we
define the grand canonical and canonical thermodynamic ensembles and
compute the associated Gibbs and Helmholtz thermodynamic potentials,
contrasting the techniques used (and results obtained) to those of our
previous work. In particular, we note that we can obtain an intrinsic
definition of these quantities in Euclidean quantum gravity, by
sidestepping some of the technical subtleties ---encountered in the
``background subtraction'' technique for regularizing the action--- in
favour of the ``counterterm subtraction'' technique\cite{pervijay,ejm}. In
the rest of the section, we examine the features of these potentials
quite closely, in preparation for later detailed studies.  In
section~\ref{sec:swallowtail}, we use the equation of state and the
first law of thermodynamics to identify the origins of the crucial
features of the shape of the Helmholtz potential (free energy). This
``swallowtail'' shape is responsible for the interesting phase
structure in the canonical ensemble. Section~\ref{sec:stability}
examines the conditions for thermodynamic stability of the black
holes, examining the specific heats and permittivity of the black
holes. In this way, we identify the stable regions of the solution
space of the equation of state. We use this stability information,
together with the information gained in earlier sections, to deduce
the refined phase diagrams exhibited in section~\ref{sec:phases}, and
some details of the phase diagrams (the slope and convexity of the
coexistence curves) are refined by using the Clayperon
equation in section~\ref{sec:coexist}.

As already stressed in this section, the thermodynamic quantities and
studies performed in those sections are rooted firmly in a microscopic
description. This is ensured by the fact that we can in principle
embed this entire discussion into a complete theory of quantum
gravity: string (and/or) M--theory. In practical terms, this
microscopic description ---the ``statistical mechanics'' underlying
the thermodynamics--- is summarized neatly in terms of the
holographically dual field theory. In this way, therefore we may carry
our calculations further and examine the nature and magnitude of the
microscopic fluctuations of the various thermodynamic quantities we
have computed, knowing that we have a description of their origin in
field theory. Thus, we find in section~\ref{sec:fluct} that the
fluctuations behave in a way consistent with the underlying
microscopic physics being supplied by the field theory: the size of
the (squared) fluctuations is controlled by a prefactor which
corresponds to precisely the inverse of the number of degrees of
freedom of the dual field theory. We observe that the size of the
fluctuations diverges as the system approaches a critical point in the
$(Q,T)$ plane.

Through most of the paper, we carried out our computations for the
four dimensional case, in order to keep many of our formulae
simple. Section~\ref{sec:higher} collects together some of the results
for the computation of various quantities. We stress that the
qualitative structure of the physics is the same for all dimensions
$d{\geq}4$, where $d{=}n{+}1$. Briefly, we also discuss in that
same section the issue of the meaning of the formal definition of
other thermodynamic ensembles by Legendre transform. It is not always
the case that the thermodynamic quantities thus defined may be arrived
at by (known) computations in Euclidean quantum gravity. Therefore,
interpretations of the physics of such ensembles are to be taken with
(at least) a pinch of salt, until such time as new technology becomes
available to compute the relevant quantities directly in quantum
gravity, as we have done here for the fixed potential (grand
canonical) and fixed charge (canonical) ensembles.

Section~\ref{sec:critical} discusses the underlying structure of the
phase structure of the canonical ensemble in the neighbourhood of the
critical point. In particular, the physics local to critical point is
universal for all of dimensions $d{\geq}4$. The critical point is a
second order phase transition point at the end of a coexistence line
of first order phase transitions. As such, it has a universal
description in terms of a Landau--Ginzburg model, with a quartic
potential ---$A_3$ in the $A$--$D$--$E$ classification of such
potentials.  The deformation of this potential gives the classic
``cusp'' catastrophe which underlies the critical behaviour, as is
well known from the van--der Waals--Maxwell description of the
liquid--gas system, with which our black hole physics shares many
features, as originally reported in ref.~\cite{cejmii}.

In closing the introduction,
we would like to stress once again how elegantly the
properties of anti--de Sitter space yield charged black hole physics
so closely akin in structure to that of ordinary field theory--like
systems, with which we have more intuition.

{}From the point of view of the Maxwell part of the action, the black
holes are nothing more than spherical capacitors, and as such, the
amount of energy they can store grows with the charge on them, but
falls with increasing hole radius. From the point of view of the
Einstein--Hilbert action however, the black holes store an amount of
energy which grows with radius. After a little thought, one might
expect on general grounds, therefore, that there might be an
interesting phase structure resulting from a competition between these
two pieces of the action.

Such reasoning on its own would not be enough to genuinely fill out
the whole $(Q,T)$ phase diagram, as the equation of state needs
additional structure. It is the presence of (negative) cosmological
constant which provides this final part: First, it provides black hole
solutions which are thermally stable in ensembles involving fixed
temperature\cite{hawkpage}, but secondly, as it defines a new length
scale, is allows the system to distinguish, on the one hand, black
holes which are large from those which are small, and on the other
hand, black holes which have small charge, from those with large
charge.

It is because of these features that the charged black hole
thermodynamics has a chance to be similar to the van--der Waals model
of the liquid--gas. Recall that without the inclusion of the effects
of the length scales set by finite particle size, on the one hand, and
attractive inter--particle forces on the other, that system would have
only the much less interesting physics of the ideal gas: there would
be no competing effects, as a function of length scale, with which to
trigger a phase transition. These basic features of AdS  give
holography a chance to work in a way which is consistent with our
intuition that the microscopic physics should be modelled by ordinary
field theory.

\section{\bf Charged AdS Black Holes}\label{sec:theholes}
For space--time dimension $n{+}1$, the
Einstein--Maxwell--anti--deSitter (EMadS$_{n+1}$) action may be
written as\footnote{We scale the gauge field $A_\mu$ so as to absorb
the prefactors involving the $U(1)$ gauge coupling into the action.}
\begin{equation}
I = -\frac{1}{16{\pi}G} \int_{M} d^{n+1}x \sqrt{-g} \left[R - F^2 +
\frac{n(n-1)}{l^2}\right]\ ,
\label{actionjackson}
\end{equation}
with ${\Lambda}{=}{-}\frac{n(n-1)}{2l^2}$ being the cosmological
constant associated with the characteristic length scale $l$.  Then
the metric on the Reissner--Nordstr\"om--anti--deSitter (RNadS)
solution may be written in static coordinates
as\cite{romans,lee,cejmii}
\begin{equation}
ds^2 = -V(r)dt^2 + \frac{dr^2}{V(r)} +r^{2}d{\Omega}^2_{n-1}\ ,
\end{equation}

\noindent where $d{\Omega}^2_{n-1}$ is the metric on the round
unit $(n{-}1)$--sphere, and the function $V(r)$ takes the form
\begin{equation}
V(r) = 1 - \frac{m}{r^{n-2}} + \frac{q^2}{r^{2n-4}} + \frac{r^2}{l^2}\ .
\end{equation}

\noindent Here, $m$ is related to the ADM mass of the hole, $M$
(appropriately generalized to geometries asymptotic to
AdS\cite{adhh}), as
\begin{equation}
M={(n-1)\omega_{n-1}\over 16\pi G}m\ ,
\end{equation}
 where $\omega_{n-1}$ is the volume of the unit $(n{-}1)$--sphere.
The parameter
$q$ yields the charge
\begin{equation}
Q=\sqrt{2(n-1)(n-2)}\left({\omega_{n-1}\over 8\pi G}\right)q\ ,
\label{thecharge}
\end{equation}
of the 
(pure electric) gauge potential, which is
\begin{equation}
A=\left(-{1\over c}{q\over r^{n-2}}+\Phi\right)dt\ ,
\label{pure}
\end{equation}
where
\begin{equation}
c=\sqrt{2(n-2)\over n-1}\ ,
\end{equation}
and $\Phi$ is a constant (to be fixed below). If $r_+$ is the largest
real  positive root of $V(r)$, then in order for this RNadS metric to
describe a charged black hole with a non--singular horizon at $r{=}r_+$, the
latter must satisfy 
\begin{equation}\label{extbound}
\left({n\over n-2}\right) r_+^{2n-2}+l^2 r_+^{2n-4} \geq q^2 l^2\ .
\end{equation}
Finally, we choose
\begin{equation}
\Phi={1\over c}{q\over r_+^{n-2}}\ ,
\end{equation}
which then fixes $A_t(r_+){=}0$, as is required by (Euclidean)
regularity of the one--form potential (\ref{pure})
at the fixed point set of the Killing vector
$\partial_t$. The physical significance of the quantity $\Phi$, which
plays an important role later, is that it is the electrostatic
potential difference between the horizon and infinity.

If the inequality in eqn.~(\ref{extbound}) is saturated,  the horizon is
degenerate and we get an extremal black hole. This inequality imposes
a bound on the black hole mass parameter of the form $m{\geq}m_e(q,l)$.

In passing to the thermodynamic discussion, we define the Euclidean
section ($t{\to}i\tau$) of the solution, and identify the period,
$\beta$, of the imaginary time with the inverse temperature. Using the
usual formula for the period, $\beta{=}{4\pi}/{V^{\prime}(r_{+})}$,
which arises from the requirement of regularity of the solution, we
obtain:
\begin{equation}
\label{betaform}
\beta = \frac{4{\pi}l^{2}r_{+}^{2n-3}}{nr_{+}^{2n-2} +
(n-2)l^{2}r_{+}^{2n-4} - (n-2)q^{2}l^{2}}\ .
\end{equation}

\noindent This may be rewritten in terms of the potential as:
\begin{equation}
\beta = \frac{4{\pi}l^{2}r_+}{(n-2)l^{2}(1-c^2\Phi^2)+nr_+^2}\ .
\label{betaformtwo}
\end{equation}

For simplicity, we will specialize to $n{=}3$ (therefore working with
EMadS$_4$) to avoid cluttering our expressions with complicated
dependences on $n$.  Our results will remain qualitatively the same
for higher $n$ (see the comments in section~\ref{sec:critical}), 
and we list some of the $n$--dependent formulae in
section~\ref{sec:higher}. Our analysis is further simplified by
adopting the following rescalings (once
we have set $n{=}3$):
\begin{equation} T \rightarrow {2\pi l\over \sqrt{3}}T\ ,
\qquad Q \rightarrow {\sqrt{3}\over l} GQ\ ,
\qquad \Phi \rightarrow \Phi\end{equation}
and for the various thermodynamic quantities used in ref.\cite{cejmii},
\begin{equation}\{W, F, E\} \rightarrow
 {\sqrt{3}\over l} G \{W, F, E\}\ ,\qquad
S\rightarrow {3G\over2\pi l^2}S\ .
\end{equation}
\beq {\rm and}\qquad r_+\rightarrow{\sqrt{3}\over l}r_+\ .
\eeq
Essentially we are introducing a system of dimensionless quantities
in which everything is measured in units of the AdS scale $l$. 
This scaling is chosen so that the thermodynamic formulae still
all have their standard form, {\it i.e.,} \beq
dE = TdS+\Phi dQ\ .
\qquad
dF=-SdT+\Phi dQ \qquad dW=-SdT-Qd\Phi\ ,\ \ {\it etc.}
\label{firstlaw}
\eeq
In the following, all of the quantities which follow are the rescaled
dimensionless quantities, unless stated otherwise.

\section{Equation of State}\label{sec:state}

\begin{figure}[hb]
\hskip1cm
\psfig{figure=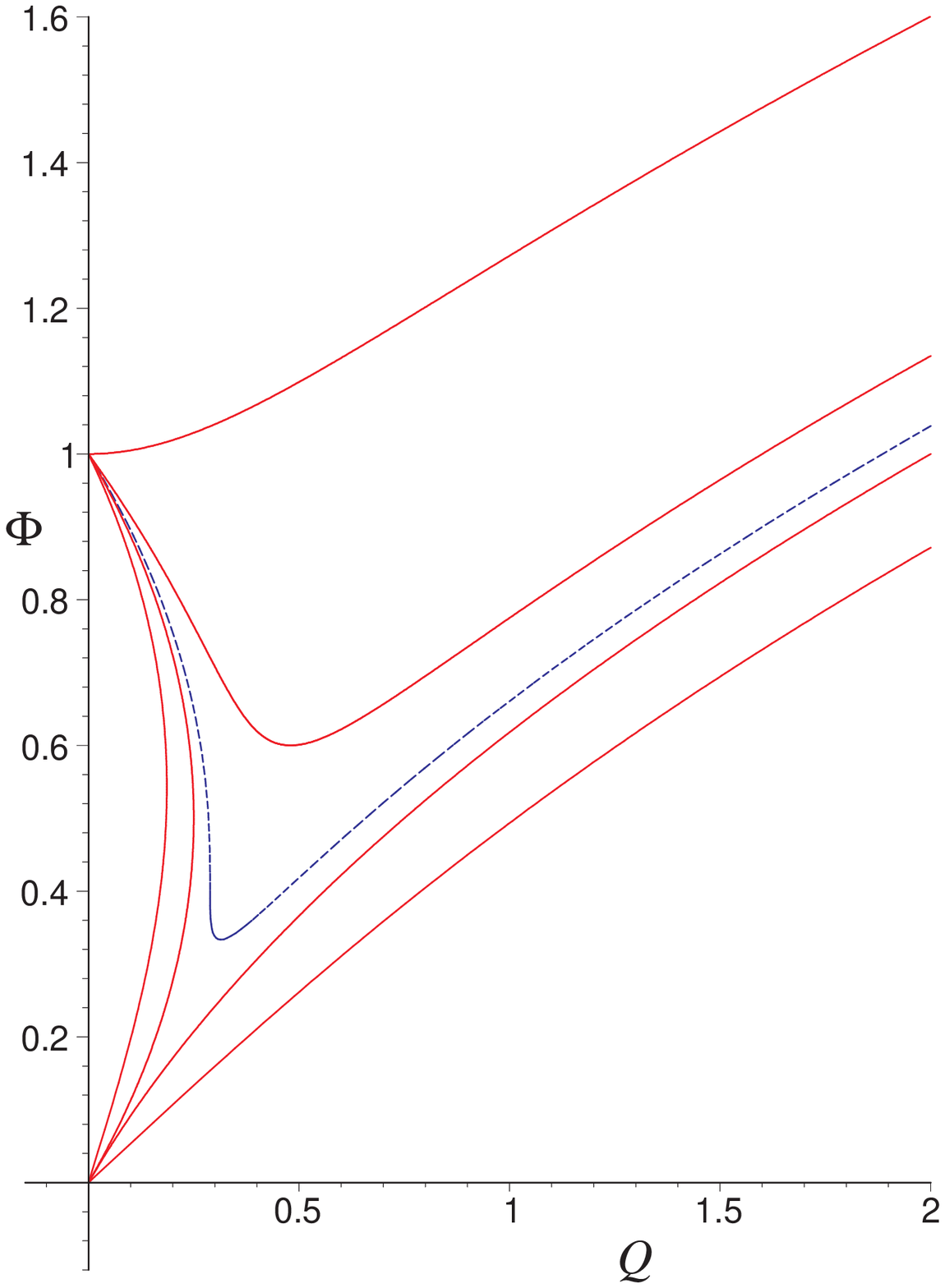,height=3.0in}
\hskip2cm
\psfig{figure=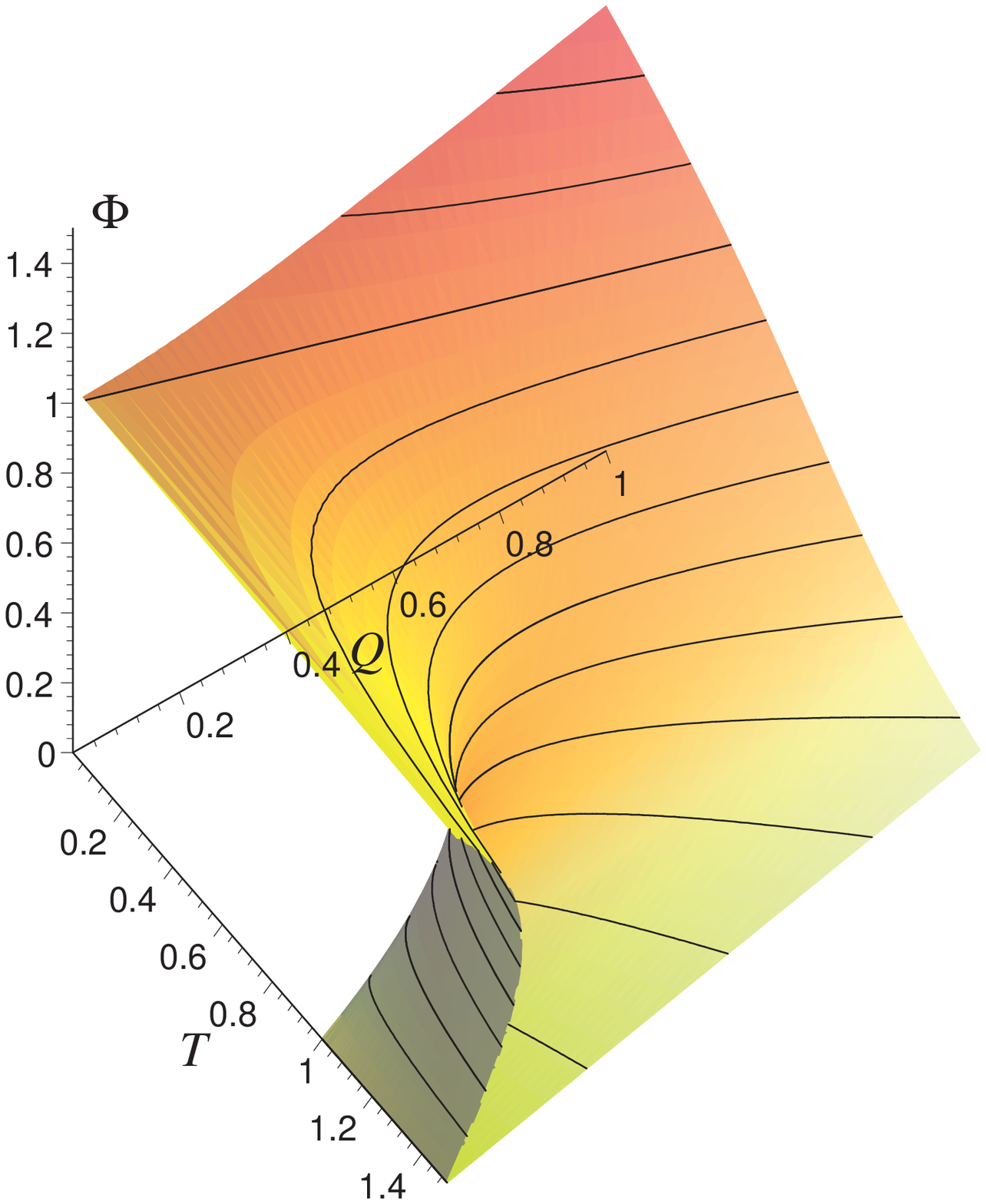,height=3.2in}
\caption{Plots of the equation of state of $\Phi$ {\it vs.}~$Q$,
showing isotherms above and below the critical temperature $T_{\rm
crit}$. For~$T{<}T_{\rm crit}$, there is only one branch of solutions,
while for $T{>}T_{\rm crit}$, there are three branches. The values of
$T$ for the isotherms plotted are (top down) $T{=}0,0.8,T_{\rm
crit},1.0,1.2$. The central (dotted) curve is at the critical
temperature.}
\label{fig:state}
\end{figure}

The Euclidean regularity at the horizon discussed at
eqn.~(\ref{betaform}) is equivalent to the condition that the black
hole is in thermodynamical equilibrium.  The resulting
equation~(\ref{betaform}) may therefore be written as an equation of
state $T{=}T(\Phi,Q)$ (analogous to $T{=}T(P,V)$ for, say, a gas at
pressure $P$ and volume $V$). For $n{=}3$, one finds:
\begin{equation}
T={\Phi^2 (1-\Phi^2) +Q^2 \over 2Q\Phi}\ .
\end{equation}
One can also solve for $Q$ as
\begin{equation}
Q= T\Phi \pm \Phi \sqrt{T^2 +\Phi^2-1}\ .
\label{eqn:state}
\end{equation}
{}From this equation of state we see that for fixed $\Phi$ we get two
branches, one for each sign, when the discriminant under the square root is
positive.  For fixed $Q$, $T(\Phi)$ has three branches for $Q{<}Q_{\rm
crit}$ and one for $Q{>}Q_{\rm crit}$, where the critical charge is determined
solving for the ``point of inflection'' where $\left(\partial
Q/\partial\Phi\right)_T{=}\left(\partial^2 Q/\partial\Phi^2\right)_T{=}0.$
In the dimensionless units used here,
one finds $Q_{\rm crit} {=}1/(2\sqrt{3})$, $T_{\rm
crit}{=}2\sqrt{2}/3$, $\Phi_{\rm crit}{=}1/\sqrt{6}$, $E_{\rm
crit}{=}\sqrt{2}/3$, and $r_{\rm +(crit)}{=}1/\sqrt{2}$. It is useful
to plot the $(\Phi,Q)$ isotherms, {\it i.e.,} plot $Q(\Phi)$ for fixed
$T$, and we exhibit these in figure~\ref{fig:state}.

As $T$ goes to zero, we approach the extremal black holes. Their
equation of state is
\begin{equation}
\Phi_e^2 ={1\over 2} (1+\sqrt{1+4 Q^2}),\qquad {\rm for\; arbitrary}\, T.
\end{equation}
For some later computations, it is often convenient to use as an
additional, non--independent parameter, the black hole radius $r_+$,
in terms of which
\begin{equation}
Q=r_+\sqrt{r_+^2-2r_+T+1},\quad 2T=r_+ +{1\over r_+}-{Q^2\over r_+^3}\ ,
\label{param1}
\end{equation}
and 
\begin{equation}
\Phi={Q\over r_+}=\sqrt{r_+^2-2r_+T+1}\ .
\label{param2}
\end{equation}

\section{The grand canonical and canonical ensembles}\label{sec:ensemble}

In thermodynamic parlance, the ``grand canonical ensemble'' is defined
by coupling the system to energy and charge reservoirs at fixed
temperature $T$ and potential $\Phi$ (an intensive variable). The
associated thermodynamic potential is the Gibbs free energy,
$W[T,\Phi]{=}E{-}TS{-}\Phi Q$. Holding the extensive variable, $Q$,
fixed, on the other hand, defines the canonical ensemble, with its
associated thermodynamic potential is the Helmholtz free energy
$F[T,Q]{=}E{-}TS$.  See section~\ref{sec:higher} for a brief
discussion of other ensembles.

In ref.~\cite{cejmii}, the calculation at fixed potential was carried
out by computing the action {\it \'{a} la} Gibbons--Hawking. With that
technique, one must regularize the computation (as the action is
formally infinite) by subtracting a contribution from a ``reference''
background which matches the solution of interest asymptotically,
giving a definition of the action relative to that of the reference
spacetime. In this case it is appropriate to use AdS ---with a fixed
(pure gauge) potential at infinity--- as the reference background.

Remarkably AdS spacetime provides for another regularization which
yields an {\it intrinsic} definition of the action. In other words,
the computation makes no reference to any other solution of the
equations of motion.  Instead, the method\cite{pervijay,ejm} proceeds by
adding a series of boundary counterterms to the action.
We refer to this as the
``counterterm subtraction'' method of defining the action, a technique
tailored to spacetimes which are locally asymptotic to anti--de
Sitter, as the counterterms are defined on the natural boundary, with
which such spaces are endowed, using the AdS scale $l$.
Also note that the inclusion of additional sectors
to the gravitational and cosmological parts of the action, such as Maxwell
terms, does not affect the definitions and therefore we can still use the
same counterterms in the present context.

\begin{figure}[hb]
\hskip5cm
\psfig{figure=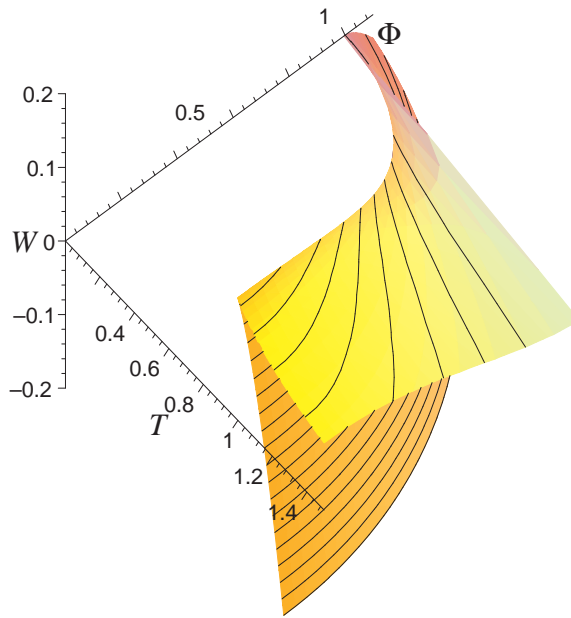,height=3.2in}
\caption{Plots of the Gibbs potential $W[\Phi,T]$ in three dimensions.}
\label{fig:gibbs3d}
\end{figure}

\begin{figure}[hb]
\hskip-0.5cm
\psfig{figure=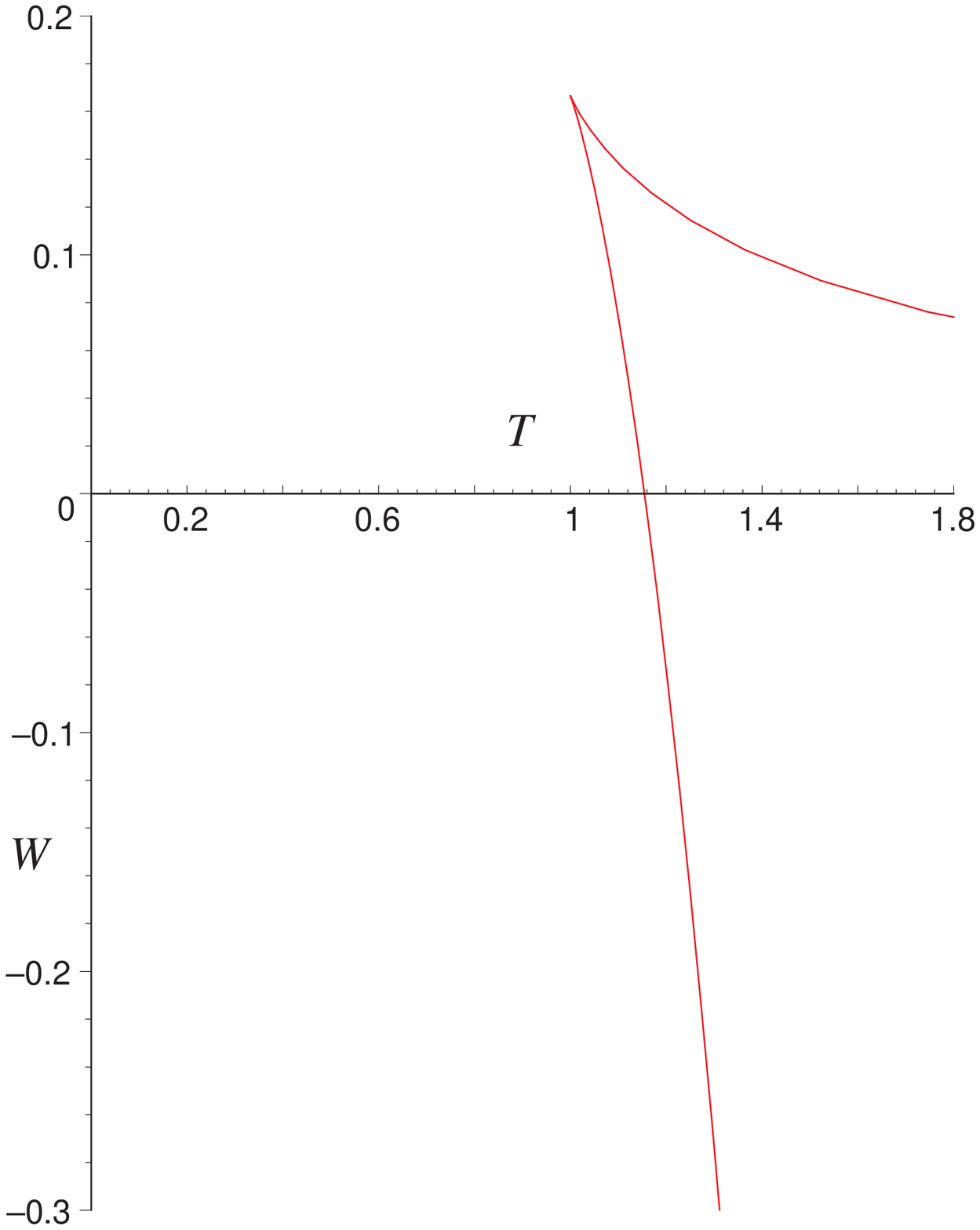,height=2.8in}
\psfig{figure=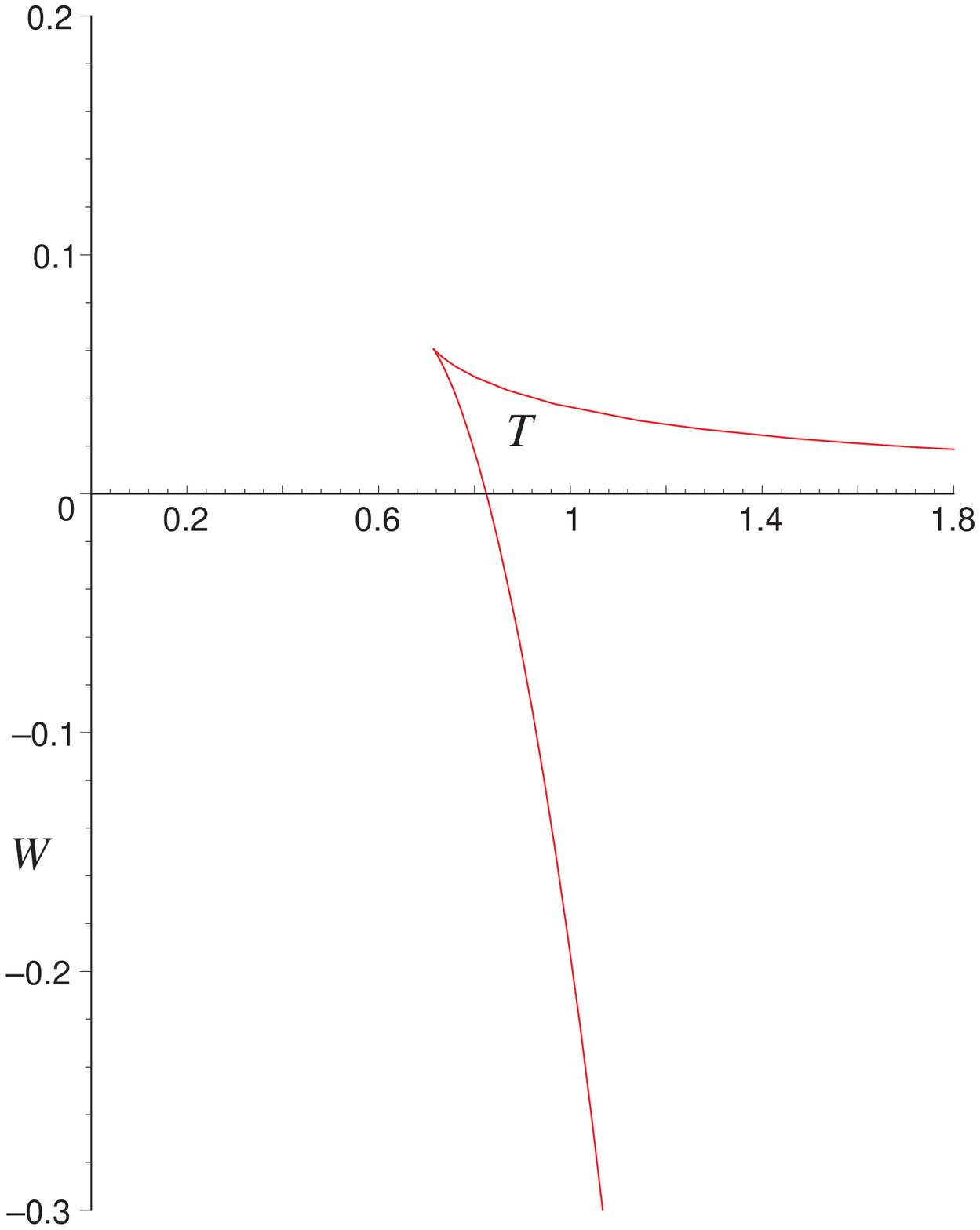,height=2.8in}
\psfig{figure=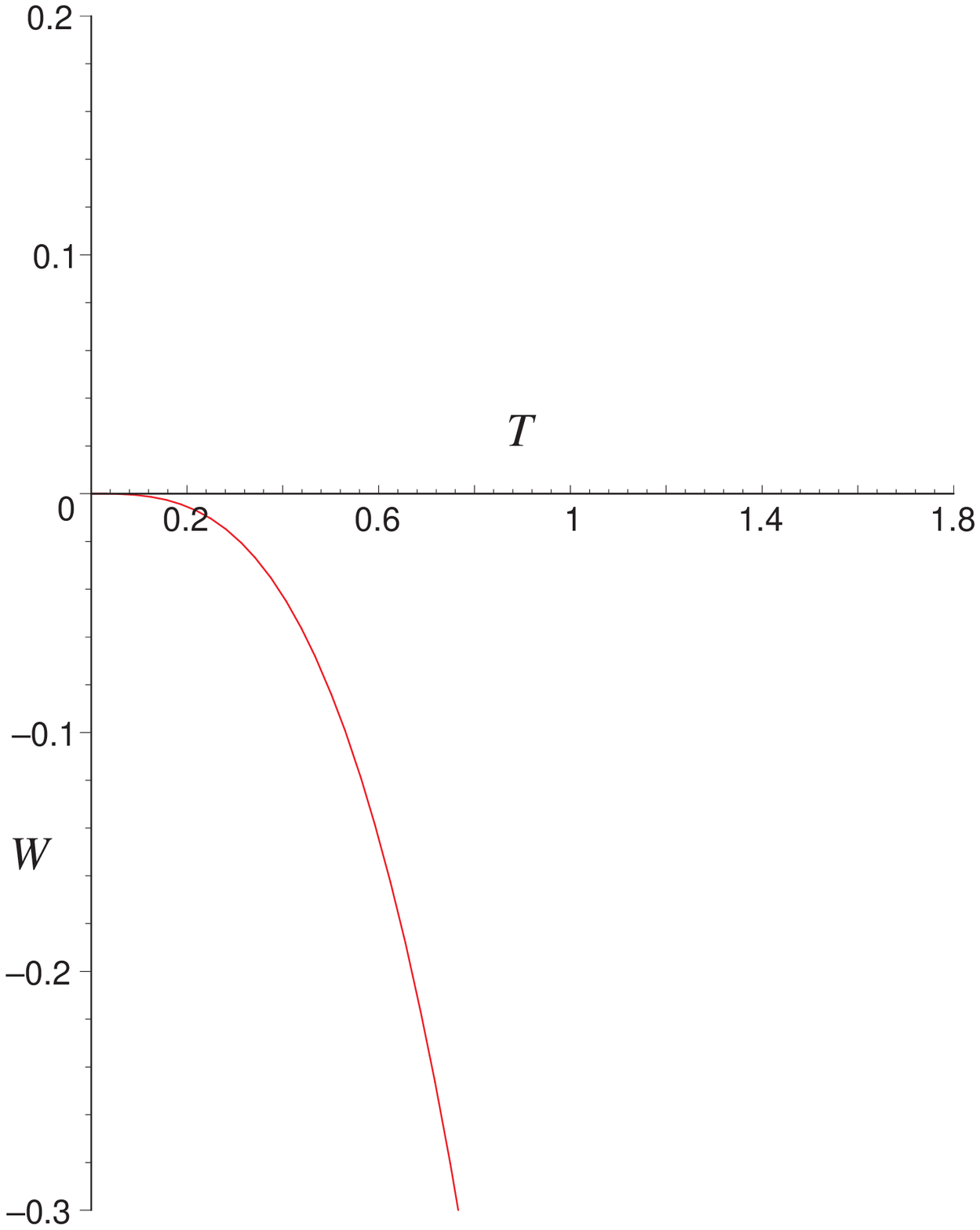,height=2.8in}
\caption{Slices of the Gibbs potential $W[\Phi,T]$, for $\Phi{=}0$, $\Phi{=}0.7$ 
and
$\Phi{=}1$.}
\label{fig:gibbssnaps}
\end{figure}

The results, using either the reference background or the counterterm
subtraction methods, are identical for the particular case in which we
want to fix the potential\footnote{For even values of $n$ there
appears a Casimir energy term \cite{pervijay}, which is immaterial for
the discussion of thermodynamics here.}, since it is possible to have
AdS space as a background solution at arbitrary temperature and
(constant) potential (but, crucially, see later).  In the present
notation, the answer is:
\begin{equation}
W[\Phi,T]={1\over 12} \left[ 3 {Q\over 
\Phi}(1-\Phi^2)-\left({Q\over\Phi}\right)^3\right]\ .
\label{eqn:gibbs}
\end{equation}
Here, $Q$ is given as $Q(\Phi,T)$ by the equation of
state~(\ref{eqn:state}). In terms of $r_+$, this is
\begin{equation}
W={1\over 12} \left[ 3 r_+(1-\Phi^2)-r_+^3\right]\ ,
\end{equation}
\noindent
and it is plotted in figure~\ref{fig:gibbs3d}, with choice slices
displayed in figure~\ref{fig:gibbssnaps}.

Turning to the canonical (fixed charge) ensemble, we wish to compute
the Helmholtz potential $F[Q,T]$ (a.k.a, the ``free energy''). In
ref.\cite{cejmii}, where we used the reference background method to
compute this, it was necessary to compute the action using an
extremal black hole as the reference background. This is because
anti--de Sitter space with a fixed charge $Q$, as measured at infinity
is {\it not} a solution of the equations of motion and so is not an
appropriate background. In order to get an intrinsic definition of
the action for fixed charge therefore, we employ the method of
counterterm subtraction, yielding:
\begin{equation}
 F[Q,T]={1\over 12} \left[ 3 {Q\over 
\Phi}-\left({Q\over\Phi}\right)^3+9Q\Phi\right]\ ,
\end{equation}
where $\Phi$ is given as $\Phi(Q,T)$ by the equation of
state~(\ref{eqn:state}).  In terms of $r_+$, $F[Q,T]$ may be written
\begin{equation}
F[Q,T]={1\over 12} \left[ 3 r_+ -r_+^3 + 9{Q^2\over r_+}\right]\ .
\label{param3}
\end{equation}

As a consistency check that we have performed the computation
correctly, note that this result may be obtained from the result for
the Gibbs potential by formally calculating the Legendre transform
$F[Q,T]{=}W[\Phi,T]{+}Q\Phi$. When computing $F$ from a Euclidean
action, the additional $Q\Phi$ term has its origin in the boundary term
introduced so as to recover the correct variational problem from the
action. It is especially satisfying to see that the counterterm
subtraction method places such intuitive relationships from
equilibrium thermodynamics on a firm footing. We shall have more to
say about this in section~\ref{sec:higher}.

In ref.\cite{cejmii}, where we computed the action using an extremal
reference background, we obtained the following expression for the free
energy (which we denote here as $\bar F$):
\begin{equation}
{\bar F}[Q,T]={1\over 12} \left[ 3 {Q\over \Phi}-\left({Q\over\Phi}\right)^3 + 
9Q\Phi-4{Q\over \Phi_e}
-8Q\Phi_e\right]\ .
\end{equation}
Note that in this case, one should consider ${\bar
\Phi}{=}\Phi{-}\Phi_e$ as the state variable, instead of $\Phi$.
Then, the first law is in this case $dE{=}TdS{+}{\bar\Phi}dQ$, and $E$
measures the energy above the extremal state. Furthermore, it is with
$\bar\Phi$ that $W[T,\Phi]$ of eqn.~(\ref{eqn:gibbs}) and ${\bar F}[T,Q]$ are 
Legendre
transforms of each other, as they should be. While $\bar F[Q,T]$ as
computed in ref.\cite{cejmii} using the extremal background is in no
way problematic, we shall not examine it further here, as the new
technology of the counterterm subtraction method has supplied us with
an intrinsic definition of the Helmholtz potential, which is the more
natural Legendre--transform partner of the Gibbs potential (\ref{eqn:gibbs}) 
found earlier. 

We shall see that the qualitative features of the results obtained in
ref.\cite{cejmii} for the canonical ensemble using $\bar F[Q,T]$ will
persist here, as the extremal background subtraction essentially
redefines the absolute normalization of some results. (The later
analysis of intrinsic stability which we do in section~\ref{sec:stability} would 
have
to be somewhat modified before direct comparison to the extremal
subtraction results, however, as we will make heavy use of the
equation of state in terms of variables $(\Phi,Q,T)$, and not the
triple $({\bar\Phi},Q,T)$ appropriate to that case.)

\begin{figure}[ht]
\hskip-0.5cm
\psfig{figure=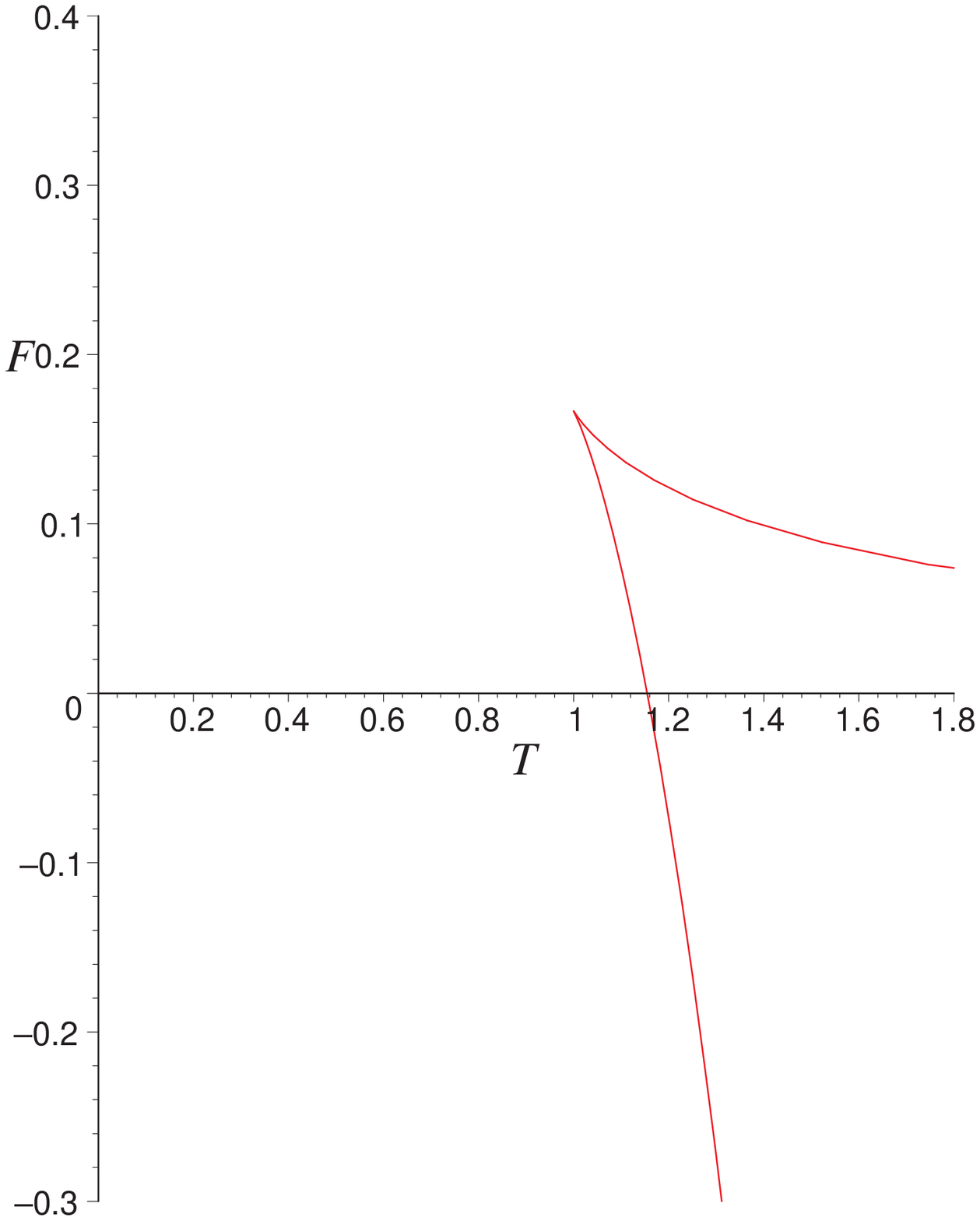,height=2.8in}
\psfig{figure=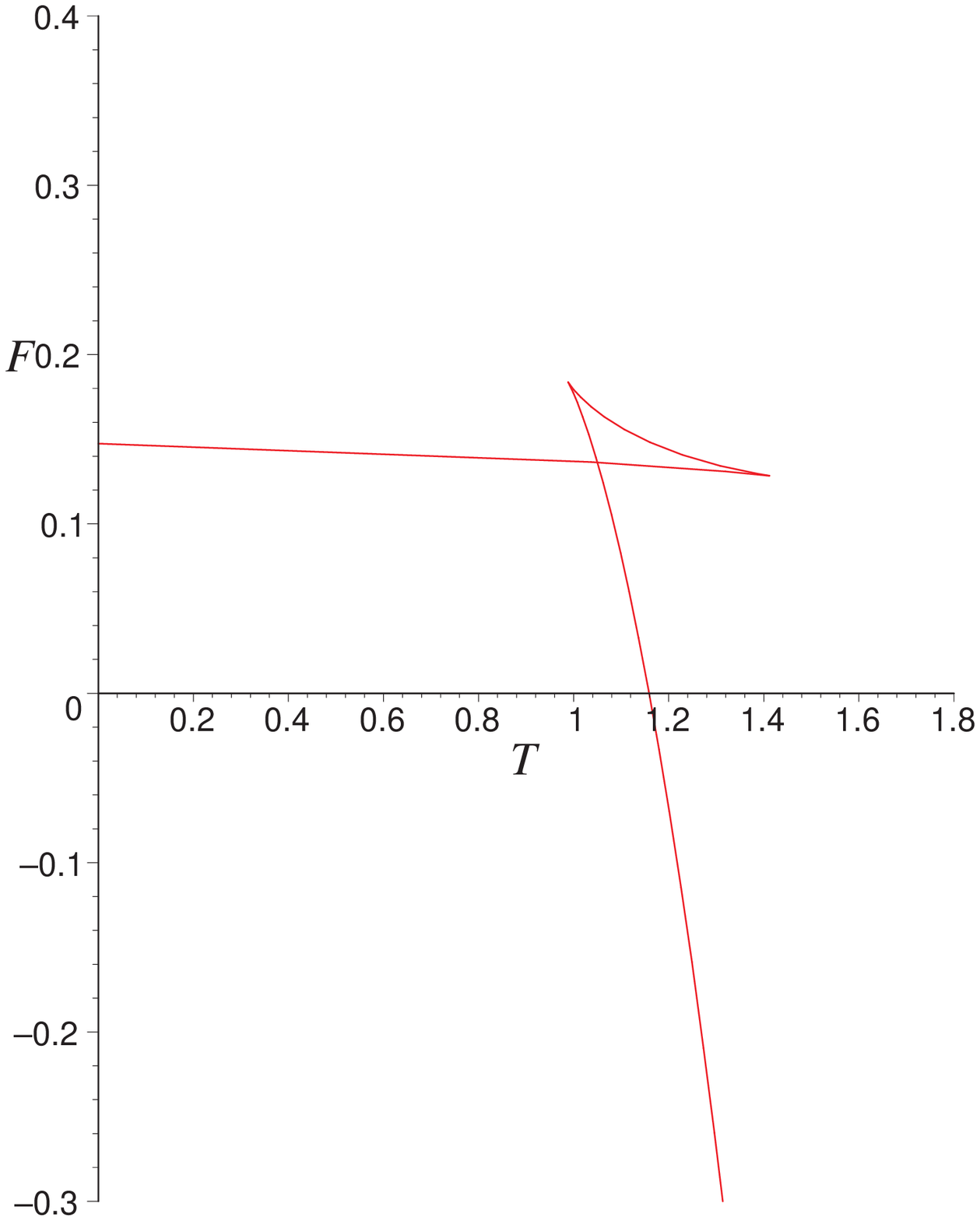,height=2.8in}
\psfig{figure=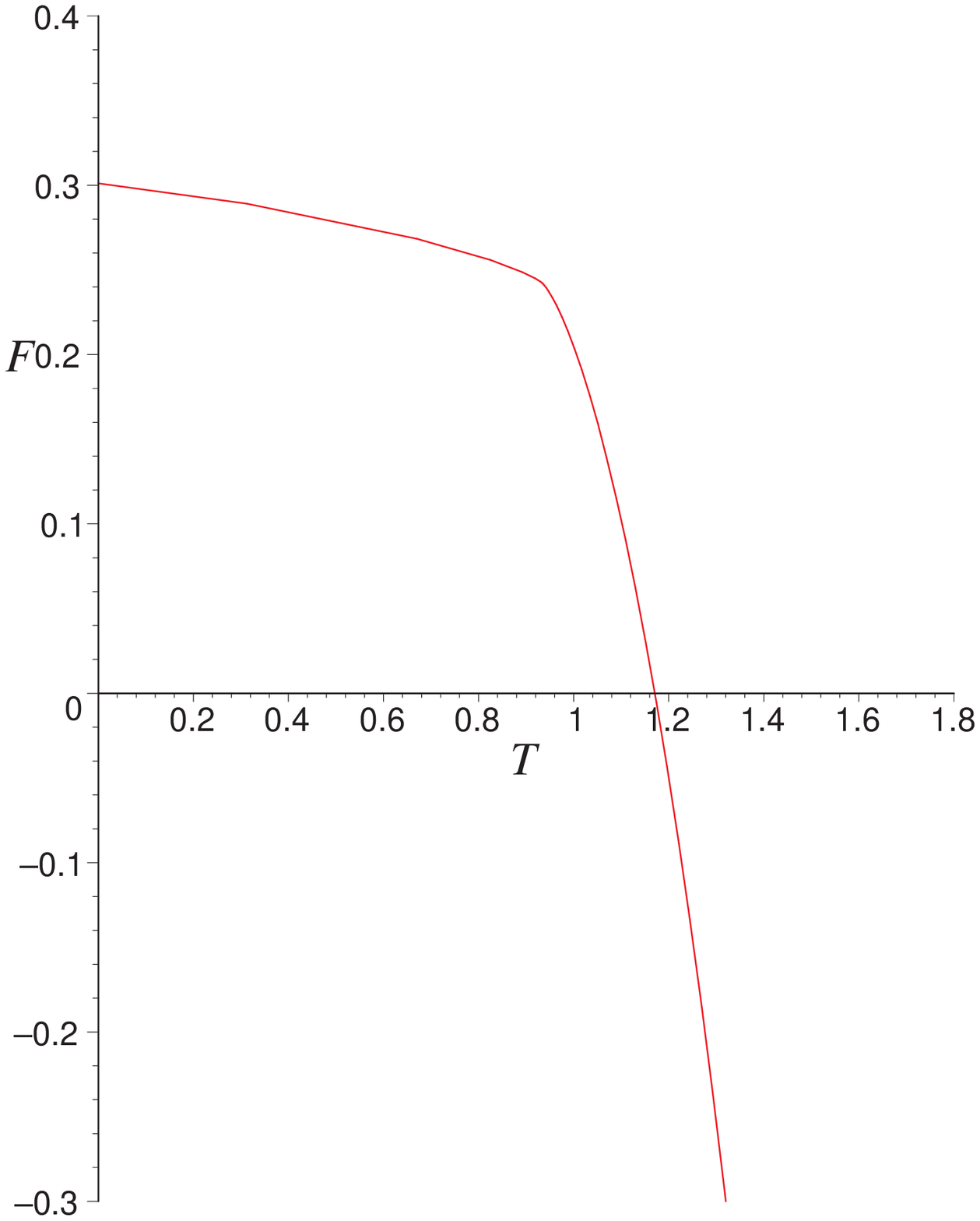,height=2.8in}
\caption{The free energy {\it vs.}~temperature for the fixed charge
ensemble, in a series of snapshots for varying charge, for values
$Q{=}0,0.15$ and $Q{=}0.299$. Note that $Q_{\rm crit}{=}0.289$, so in
the last plot, the bend (near $T_{\rm crit}{=}0.943$, is in the
neighbourhood of the critical point of second order.}
\label{fig:Tfreesnaps}
\end{figure}

We now return to the analysis of the intrinsically defined Helmholtz
potential $F[Q,T]$. It was noticed in ref.~\cite{cejmii} that a plot
of ${\bar F}(T)$ for various values of $Q$ reveals (below a $Q_{\rm
crit}$) a section of a``swallowtail'' shape, which controls much of
the phase structure (in the canonical ensemble)
 discussed there, and to be discussed later
here. (See figs.~5 and~6 of ref.~\cite{cejmii} and associated text for
details.)  The same may be observed here for $F(T)$ for varying $Q$, as
shown in figure~\ref{fig:Tfreesnaps}.

\begin{figure}[ht]
\hskip-0.5cm
\psfig{figure=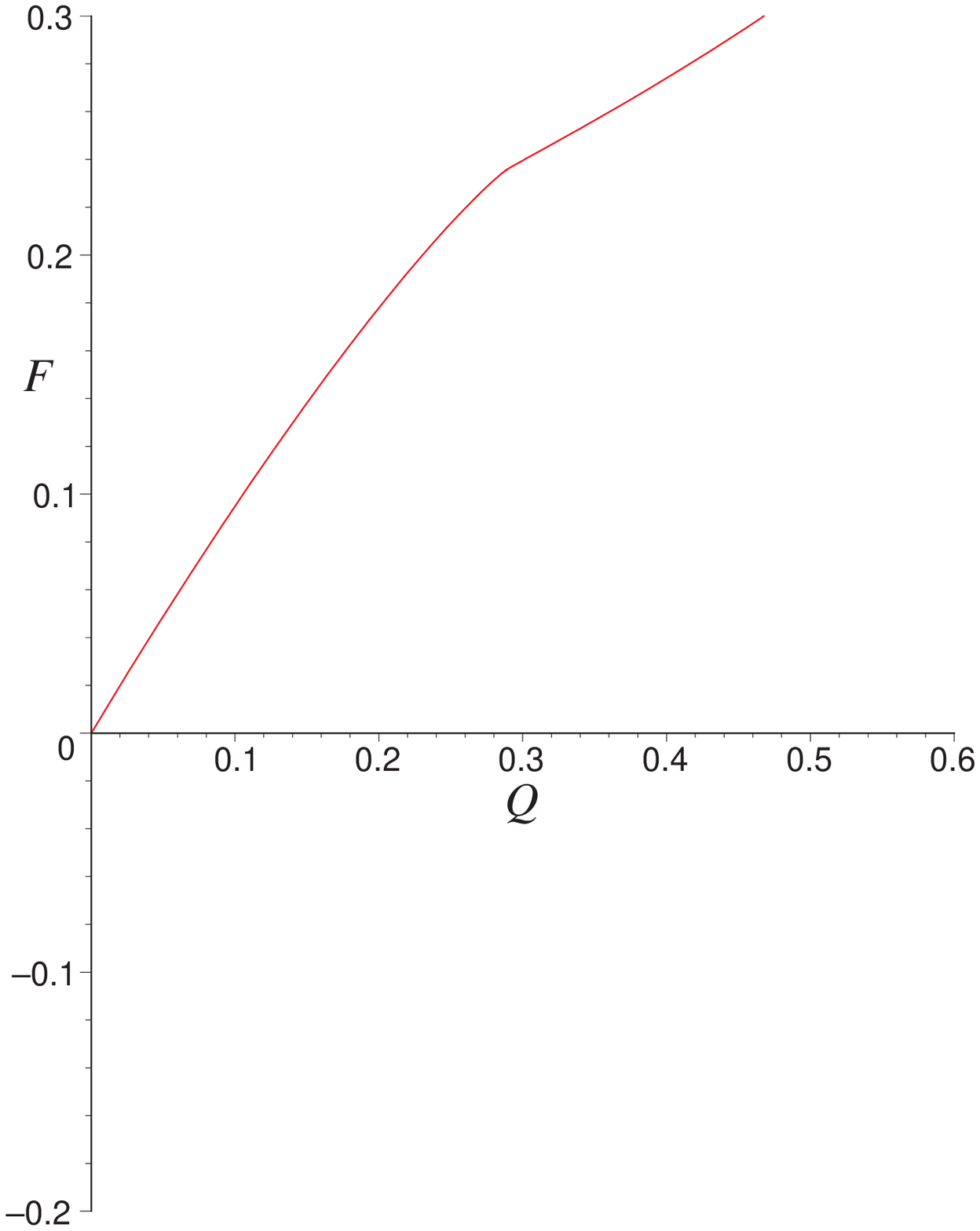,height=2.8in}
\psfig{figure=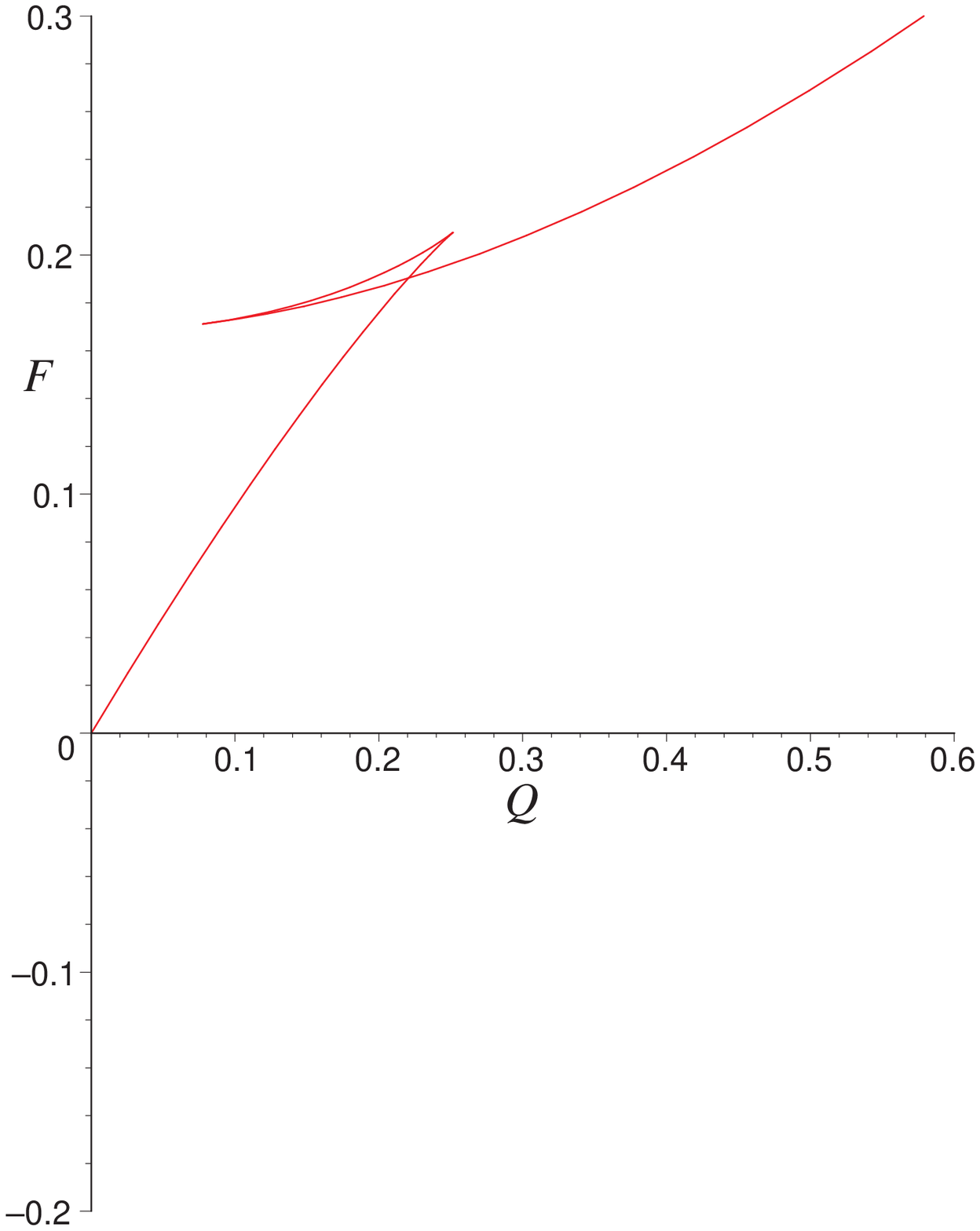,height=2.8in}
\psfig{figure=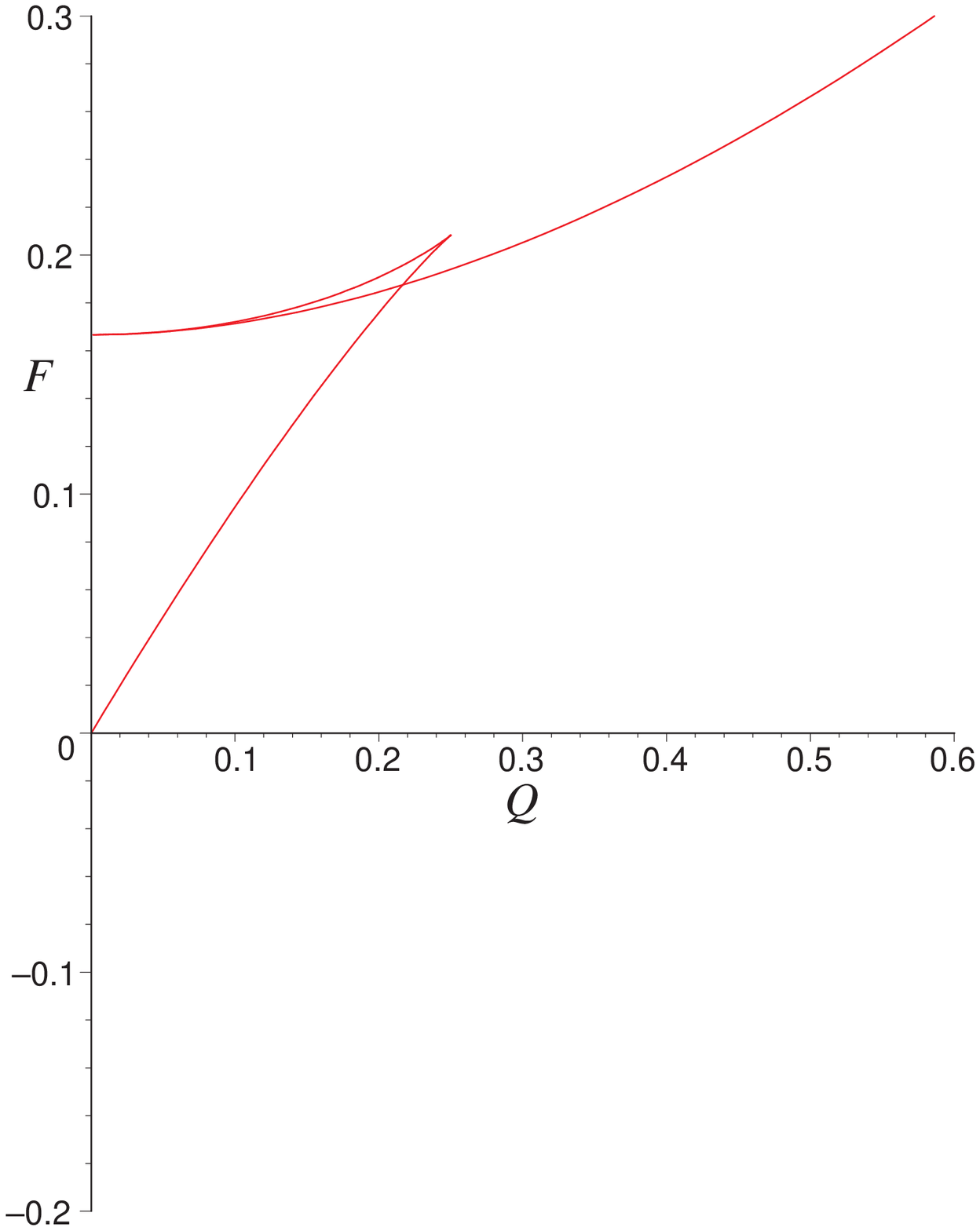,height=2.8in}
\psfig{figure=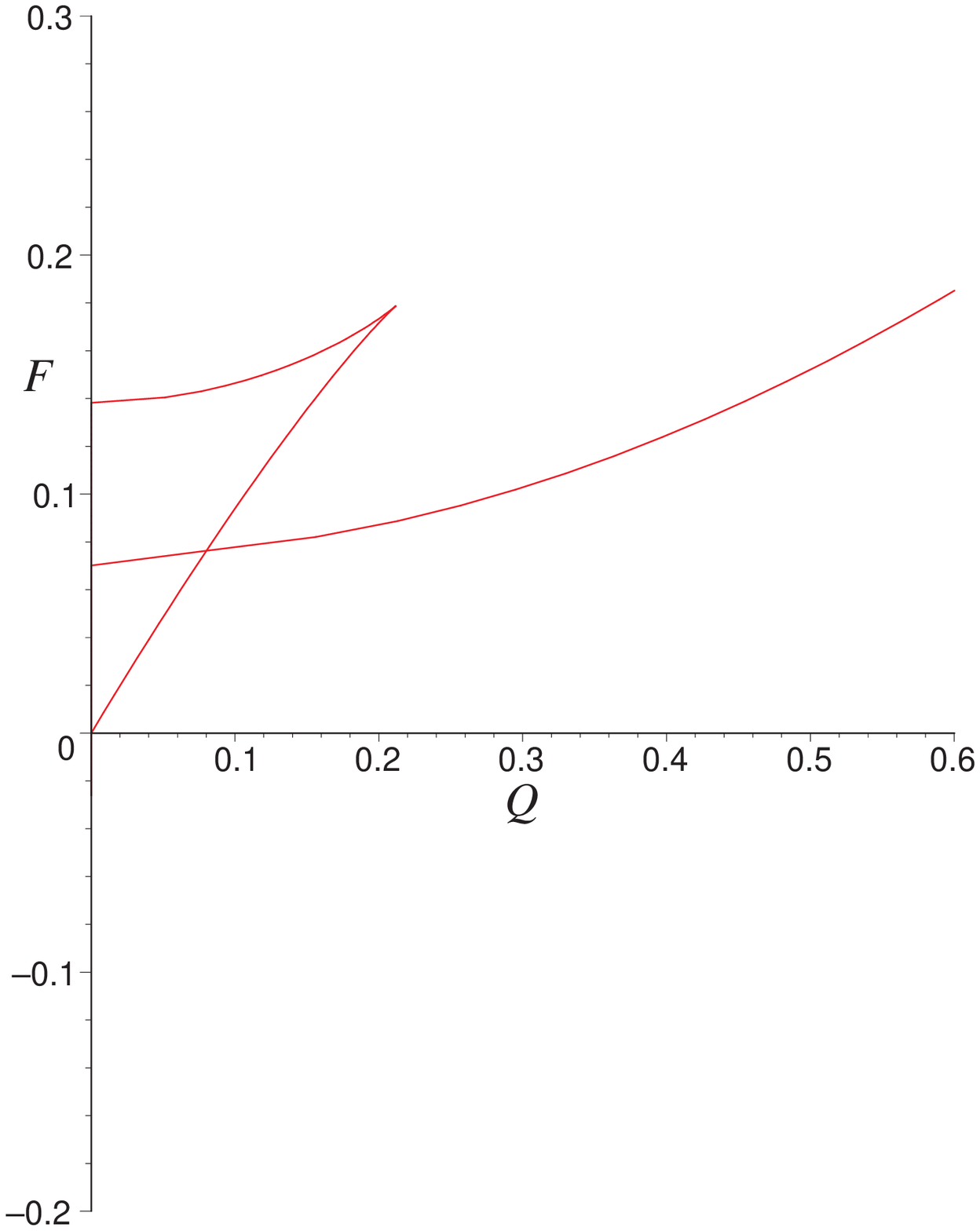,height=2.8in}
\psfig{figure=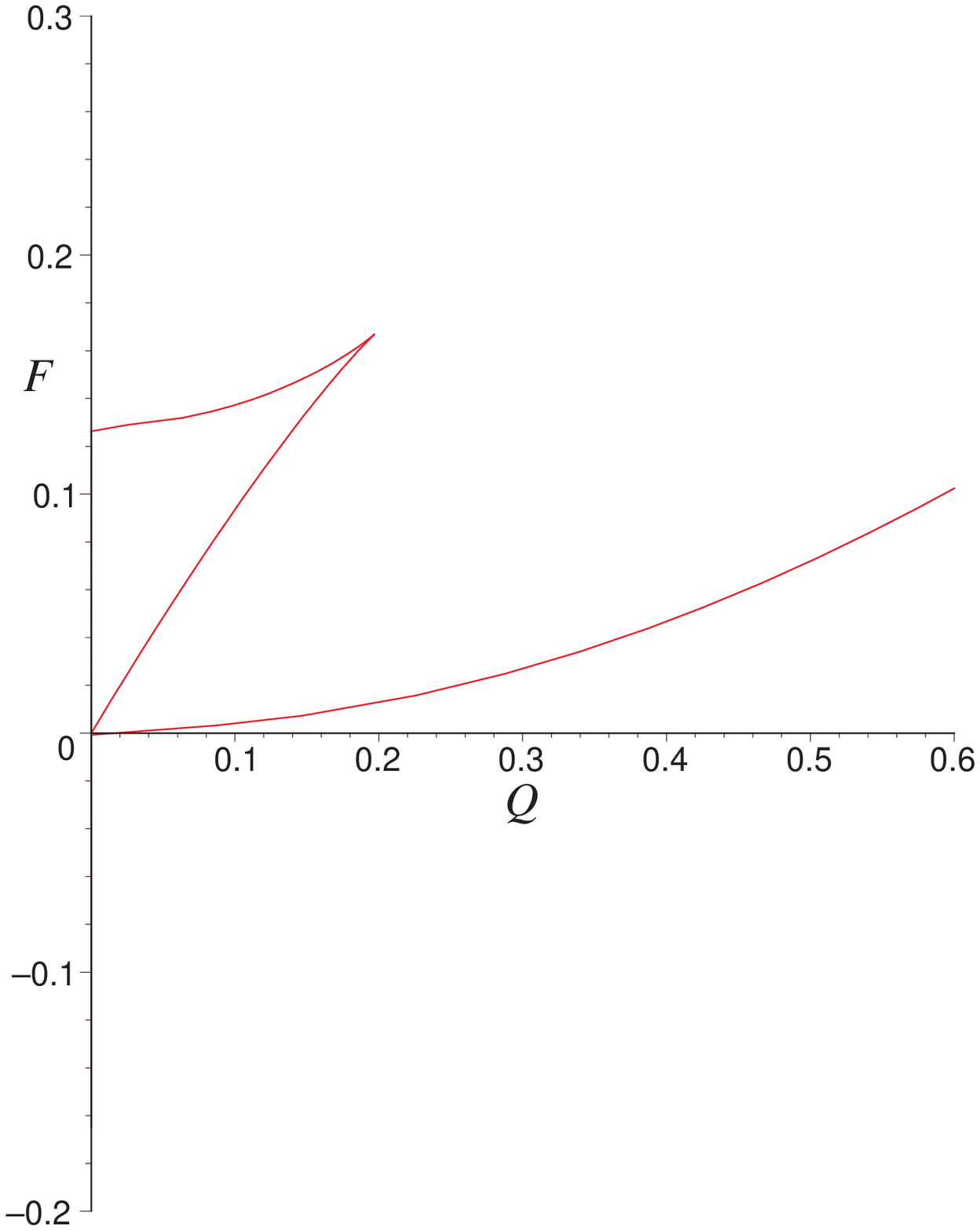,height=2.8in}
\psfig{figure=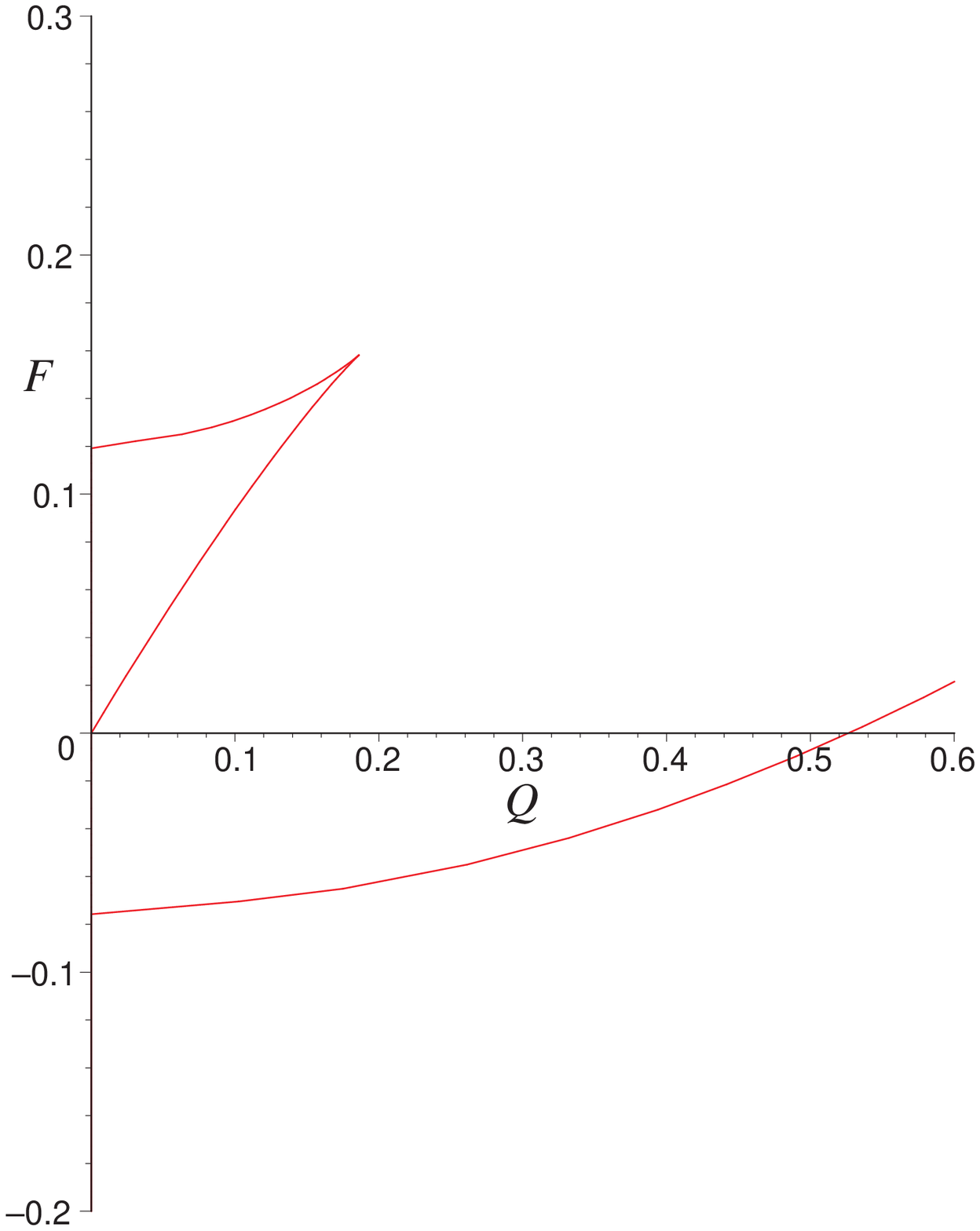,height=2.8in}
\caption{The free energy {\it vs.}~charge for the fixed charge
ensemble, in a series of snapshots for varying temperature, for values
$T{=}0.943,0.997,1.00,1.10$, and (for the ``Zorro'' plot) $T{=}T_{\rm
HP}{=}1.154$, and finally $T{=}1.20$. Note that $T_{\rm
crit}{=}0.943$, and so in the first plot, the bend (near $Q_{\rm
crit}{=}0.289$), is in the neighbourhood of the critical point of
second order.}
\label{fig:Qfreesnaps}
\end{figure}

It may be further observed that a plot of $F(Q)$ for fixed $T$ 
reveals (above a $T_{\rm crit}$), a similar swallowtail section, as
shown in figure~\ref{fig:Qfreesnaps}.

\begin{figure}[hb]
\hskip5cm
\psfig{figure=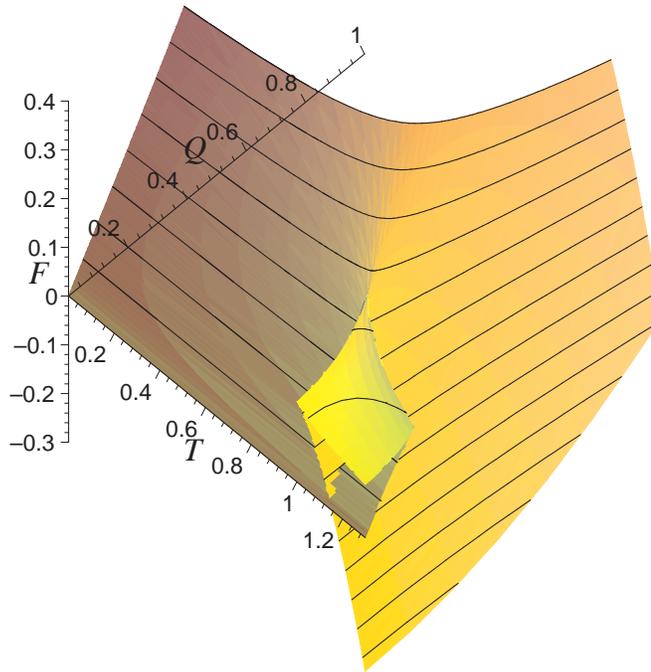,height=3.5in}
\caption{Plots of the Helmholtz potential $F[Q,T]$, in three
dimensions, clearly showing the swallowtail shape for $T{>}T_{\rm  crit}$
and $Q{<}Q_{\rm crit}$.}
\label{fig:free3d}
\end{figure}

The full three dimensional shape of $F[Q,T]$ is plotted in
figure~\ref{fig:free3d}.

That such a shape appears in the thermodynamics (above $T_{\rm crit}$
or below a $Q_{\rm crit}$) can be shown to follow from the first law
of thermodynamics, the definition of the thermodynamic potentials, and
the form of the particular equations of state which the black holes
obey. We will show how this comes about next.

\section{The swallow tales}\label{sec:swallowtail} 
The sections of swallowtails in the $F(Q)$ and $F(T)$ plots (above a
$T_{\rm crit}$ or below a $Q_{\rm crit}$) can be seen to come from the
existence of the previously mentioned 
three branches of solutions to the equation of state. We have from the first 
law, 
and the definition of the thermodynamic
potential, that $dF{=}{-}SdT{+}\Phi dQ$. Therefore, for fixed
$T$ we find
\begin{equation}
F(T)=\int\Phi(Q)\,dQ + f(T)\ ,
\label{fixedt}
\end{equation}
where $f(T)$ is an arbitrary function of $T$. The integral function
can be obtained by looking at the plot of isotherms. When we have
three branches ({\it i.e.,} $T{>}T_{\rm crit}$), the curve $\Phi(Q)$
winds back and forth in a way that the integral describes a shape with
three connected branches, constituting a section of the
``swallowtail'' shape. This can be seen by examination of the plots of
the equation of state in figure~\ref{fig:state} and the plots of the
slices $F(T)$ displayed in figure~\ref{fig:Tfreesnaps}\footnote{As
visual differentiation is often easier to perform than integration, we
gently remind the reader that the defining relation
$\Phi{\equiv}(\partial F/\partial Q)_T$ may be of use here, in
conjunction with the snapshots of $F(Q)$ for fixed $T$ given in
figure~\ref{fig:Qfreesnaps}.}.

\begin{figure}[ht]
\hskip3.0cm
\psfig{figure=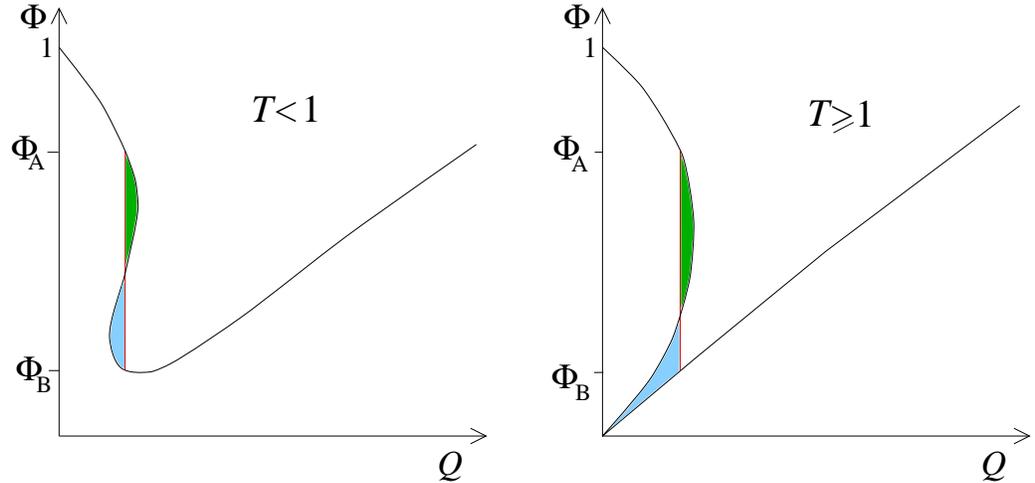,height=2.5in}
\caption{The figure on the left shows how the condition for a phase
transition may be interpreted in terms of an ``equal area law''
analogous to that due to Maxwell for the van--der Waals liquid--gas
model. For $T{\geq}1$, though, the isotherms have a very different
qualitative structure. The equal area law one might formulate,
deducing a phase transition for arbitrarily high $T{>}1$, for small
enough $Q$, actually is incorrect. See text for the resolution of the
puzzle.}
\label{fig:arealaw}
\end{figure}

Equation~(\ref{fixedt}) is usually employed to formulate an ``equal
area law" governing the phase transitions of the system.  The latter
occur at the point where the free energies of two branches, (say A and
B), are equal: $F_{\rm A}{=}F_{\rm B}$. From eqn.~(\ref{fixedt}) this
equality may be translated
into a statement about the equality of the areas enclosed by the
isotherm curves and a line of constant $Q$ in the~$(\Phi,Q)$ plane, as
shown in the sketch on the left in figure~\ref{fig:arealaw}.  There is a
subtlety, though, in using eqn.~(\ref{fixedt}) with the isotherm curves
of eqn.~(\ref{eqn:state}) for $T{>}1$ (recall that isotherms with
$T{\geq}1$ go through the origin $\Phi{=}Q{=}0$. See
figure~\ref{fig:state}.). Given that the transition is governed by the
equal area law, it would seem from the curves on the right in
figure~\ref{fig:arealaw} and the area law deduced from
eqn.~(\ref{fixedt}) that even for $T{>}1$, for which a minimum value of
$\Phi$ ceases to exist, one can always find a phase transition point
for arbitrarily large temperatures and small enough charge.  This must
be wrong since it contradicts what we know about the phase transition
from the curves of $F$ for constant $Q$, namely, that the phase
transition takes place at a temperature that is smaller (or equal, at
$Q{=}0$) than the Hawking--Page temperature $T_{\rm HP}$. (See
ref.\cite{cejmii} and the upcoming section~\ref{sec:phases} for
detailed discussion of the phase structure.)

The resolution of this puzzle is instructive, and is made manifest
most clearly by working in terms of the parameter~$r_+$, using
eqns.~(\ref{param1}),~(\ref{param2}) and~(\ref{param3}). We can
explicitly compute $F$ using eqn.~(\ref{fixedt}) as
\begin{equation}
F=\int_0^{r_+} \Phi(\bar r_+) (dQ/d\bar r_+) d\bar r_+
\end{equation}
to recover precisely eqn.~(\ref{param3}).  So, what is the reason that
``naive integration'' using the ``the equal area law'' yields a
different result?

The point is that for $T{>}1$ the function $F(Q)$ is discontinuous at
$Q{=}0$, where branches 2 and 3 separate (see for example, the last
plot of figure~\ref{fig:Qfreesnaps}). For those isotherms, there is a
range of values for $r_+$,
$T{-}\sqrt{T^2{-}1}{<}r_+{<}T+\sqrt{T^2{-}1}$ for which $Q$ and $\Phi$
become imaginary. Nonetheless, the product $\Phi dQ$ is real
throughout, and so is $F$. Then, $F(Q)$ would be a continuous function
if we plotted it in the complex $Q$ plane. In performing the
integration above for $T{>}1$ we have implicitly included the points
where $\Phi$ and $Q$ are imaginary. Notice that it is by including
these points that we recover sensible physics, since we want the
critical line to end at $Q{=}0$ at the point of the Hawking--Page
phase transition. The ``equal area law'', as it is, fails in this
instance.

\begin{figure}[hb]
\vskip-1.5cm
\hskip3cm
\psfig{figure=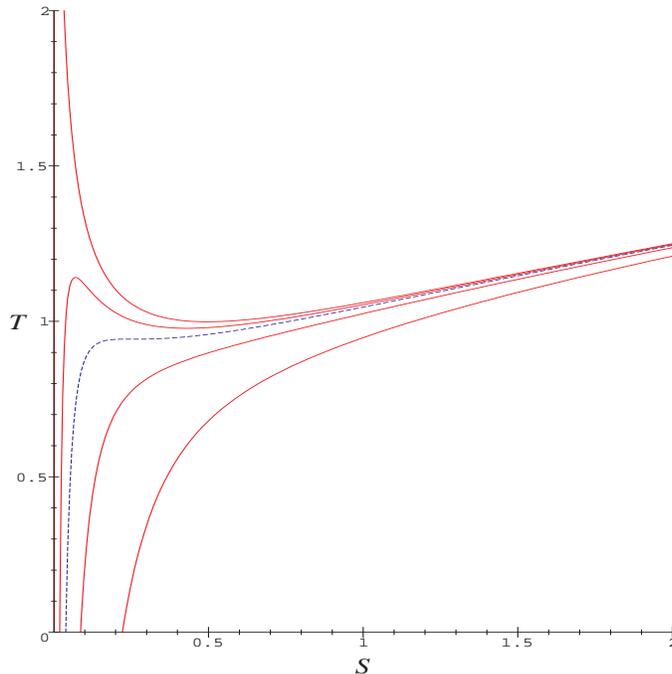,height=3.5in,width=3.5in}
\caption{Plots of the equation of state of $S$ {\it vs.}~$T$, showing
isocharge lines above and below the critical charge $Q_{\rm
crit}$. For $Q{>}Q_{\rm crit}$, there is only one branch of solutions,
while for $Q{<}Q_{\rm crit}$, there are three branches. The values of
$Q$ for the isocharge curves plotted are (top down) $Q{=}0,0.20,Q_{\rm
crit},0.45,0.80$. (The uppermost curve shows the $Q{=}0$ case, which
has two branches. The central (dotted) curve is at the critical
charge.}
\label{fig:stateS}
\end{figure}

Let us now turn to the study of the free energy for fixed $Q$. We have
\begin{equation}
F(Q)=-\int S(T) dT + g(Q)
\end{equation}
In this case we need $S(T)$. Since
\begin{equation}
S={1\over2} \left({Q\over \Phi}\right)^2= {r_+^2\over2}
\end{equation}
we can use the equation of state to plot $S(T)$ for fixed $Q$, which
is shown in figure~\ref{fig:stateS}. 
It can be readily seen that for $Q{<}Q_{crit}$ we get three branches
(notice that the qualitative features of the plot of~$S(T)$ follow
from those of $r_+(T)$ or $r_+(\beta)$ plotted in figure~3 of
ref.\cite{cejmii}, where it has a resemblance to the van der Waals
$P(V)$ curve). A section of the swallowtail again
follows\footnote{Again, one can use figure~\ref{fig:stateS} to
reconstruct $F(Q)$ visually using the integral relation, or one may
use the definition of the entropy $S{\equiv}(\partial F/\partial T)_Q$
to reconstruct figure~\ref{fig:stateS} from
figure~\ref{fig:Tfreesnaps}.}.

The astute reader may wonder why the swallowtail shape (and the
resulting liquid--gas--like) phase diagram occurs in the canonical
ensemble, where in addition to $T$, the extensive variable $Q$ is an
external control parameter, and not in the grand canonical ensemble,
where the intensive variable $\Phi$ would be the control. This is of
course what happens in the case of the van--der--Waals--Maxwell
system, where the phase diagram is in $(P,T)$ space, and not $(V,T)$
space\cite{van,stanley}.  The swallowtail shapes occur there in the
Gibbs potential. It is now hopefully clear that the answer follows
from the fact that our equation of state yields three branches of
solutions for the intensive variable $\Phi$ (or T) as a function of
the fixed extensive variable $Q$ (or $S$), as can be seen by examining
the curves displayed in figs.~\ref{fig:state} and~\ref{fig:stateS}.

That there are no swallowtail shapes in any of the other
ensembles follows from the fact that no more than two branches  occur
for the equation of state written in terms of other variables.

\section{intrinsic stability}\label{sec:stability} 
Given that we have the full power of the thermodynamic framework at
our disposal, (thanks to the stabilizing influence of a negative
cosmological constant), it is interesting to consider the
thermodynamic stability of our various solutions against microscopic
fluctuations\footnote{See refs.\cite{jorma,SandM} for analyses which
overlap with those presented here, in a similar context.}. Notice that
one can always formally compute the relevant macroscopic quantities
(like specific heats, {\it etc.,}) which we discuss here, without any
reference to an underlying microscopic description. This has been done
in the context of black hole thermodynamics since time
immemorial. {\it The difference here is that we know the nature of the
microscopic degrees of freedom which supply the underlying
``Statistical Mechanics'' which gives rise to these macroscopic
thermodynamics quantities.} The underlying physics is that of the
gauge theory to which this system is holographically dual, which in
turn is the physics of coincident branes. This will become more
apparent in section~\ref{sec:fluct} when we explicitly study the
fluctuations themselves.

Thermodynamic stability may be phrased in many different
ways\cite{callen,landau}, depending on which thermodynamic function we
choose to use, and how obscure we are attempting to seem.  For
example, it can be seen as minimization of the energy, $E$, as a
function of $(S,Q)$, or maximization of the entropy $S$, as a function
of $(E,Q)$, {\it etc}. In any case, one is considering an
infinitesimal variation of the state function away from
equilibrium. The first law (\ref{firstlaw}) will ensure that the first
order terms vanish. Stability is then a statement about the second
order variations.  Generally then the stability conditions are phrased
in terms of the restriction that the Hessian of the state function is
positive (or negative, depending on the context) semi--definite.

An equivalent but physically more transparent way of writing the
stability conditions is in terms of specific heats and other
``compressibilities'', to wit:
\begin{equation}
C_Q\equiv T\left({\partial S\over\partial T}\right)_Q \geq 0,\quad
C_\Phi\equiv T\left( {\partial S\over\partial T}\right)_\Phi \geq
0,\quad \varepsilon_T\equiv\left({\partial Q\over \partial
\Phi}\right)_T\geq 0\ .
\label{stabcrit}
\end{equation}
The first two, the specific heats at constant electric charge and
potential, are familiar analogues of the specific heats at constant
volume and pressure in fluid systems.  In the case in hand, they
determine the thermal stability of the black holes, indicating whether
a thermal fluctuation results in an increase or decrease in the size
of the black hole. (This follows from the fact that the entropy is
proportional to the size of the black hole,.) Stability follows from
$C{\geq}0$, given the fact that black holes radiate at higher
temperatures when they are smaller.  

The last quantity, $\varepsilon_T$, has the following physical
interpretation.  It is negative if the black hole is electrically unstable
to electrical fluctuations (if they are possible, see later
discussion). This happens if the potential of the black hole {\it
decreases} as a result of placing more charge on it. The potential
should of course {\it increase}, in an attempt to make it harder to
move the system from equilibrium\footnote{This follows from common
sense, or more formally, Le Chatelier's principle.}. $\varepsilon_T$
therefore deserves to be called the ``isothermal (relative)
permittivity'' of the black hole.

There are of course other interesting ``response functions'' for the
system, such as the adiabatic permittivity, $(\partial
Q/\partial\Phi)_S$ or the quantity analogous to the coefficient of
thermal expansion in liquid--gas systems, $\alpha_\Phi{=}(\partial
Q/\partial T)_\Phi$, which are not all independent. The ones which we
have discussed above will suffice for the physics that we study in
this paper.

We may examine the plots of the isotherms in figure \ref{fig:state}
and deduce that the negatively sloped branches are electrically
unstable if there are electrical fluctuations possible. Similarly, we
may deduce that the negatively sloped branches of the $(S,T)$
isocharge curves in figure~\ref{fig:stateS}, are thermally unstable,
and so on.

Stability follows, equivalently, from the concavity/convexity of the
plots of $F$ and $W$ as functions of $T$. In fact, the specific heat
conditions are equivalent to
\begin{equation} 
\left({\partial^2 F\over \partial T^2}\right)_Q\leq 0\quad {\rm
and}\quad\left({\partial^2 W\over \partial T^2}\right)_\Phi\leq 0\ ,
\label{condition1}
\end{equation} 
whereas the permittivity condition is
\begin{equation} 
\left({\partial^2 F\over \partial Q^2}\right)_T\geq 0, \quad {\rm or}\quad 
\left({\partial^2 W\over \partial \Phi^2}\right)_T\leq 0\ .
\label{condition2}
\end{equation}

By examining the isotherms displayed in figure~\ref{fig:state}, we see
that there are a number of features in the $(Q,T)$ plane which govern
electrical stability. Generically, let us describe the three branches
of an isotherm as follows: We call ``branch 3'' the branch of
solutions which extends all the way from $Q{=}\infty$, terminating
where $dQ/d\Phi{=}0$. From there, ``branch 2'' takes over, terminating
where again $dQ/d\Phi{=}0$. The isotherm continues with ``branch 1''
until the point $Q{=}0,\Phi{=}1$ is reached. This terminology matches
that of ref.\cite{cejmii}. 

{}From this definition, then, branch 3 is electrically stable for most
of its extent, except for a small region near the join with branch
2. In this case, before reaching the point where $dQ/d\Phi{=}0$ the
permittivity changes sign at a point where $dQ/d\Phi{=}\infty$ and
renders branch 3 electrically unstable thereafter. This is a feature
that is absent from the standard van--der--Waals--Maxwell system (in the
latter there are no points in the isotherms where $dP/dV=\infty$), and
which will introduce a significant modification of the phase diagram.

Branch 2, being between two places where $dQ/d\Phi{=}0$, has positive
definite slope and hence is electrically stable everywhere, while
branch~1 is electrically unstable everywhere, having negative definite
slope. To compute precisely where the electrical instability begins,
we need only find the location of the minimum of the isotherms, that
is, the above mentioned point where $dQ/d\Phi{=}\infty$, which is given
by the equation $Q{=}T\sqrt{1{-}T^2}$. With the segment of the
$T$--axis from 0 to 1, this forms a region in the $(Q,T)$ plane within
which branch~3 and branch~1 are unstable to electric
fluctuations. Branch~2 is electrically stable everywhere, as mentioned
before, but as already pointed out in ref.\cite{cejmii}, and as a
quick examination of the figure~\ref{fig:stateS} of the isocharge
$(S,T)$ curves reveals, branch~2 is unstable to thermal fluctuations,
and so never plays a role in the canonical and grand canonical
ensembles.

It is also entertaining to subject by eye the snapshots of $F$ and $W$
taken in figures~\ref{fig:gibbssnaps},~\ref{fig:Tfreesnaps}
and~\ref{fig:Qfreesnaps} to the convexity and concavity
conditions~(\ref{condition1}) and~(\ref{condition2}). We find that the
shapes of $F$ and $W$ do indeed confirm our conclusions about the
stability of the various branches.

It is very instructive to plot the boundaries of the various branches
in the $(Q,T)$ plane:

\begin{figure}[hb]
\hskip2cm
\psfig{figure=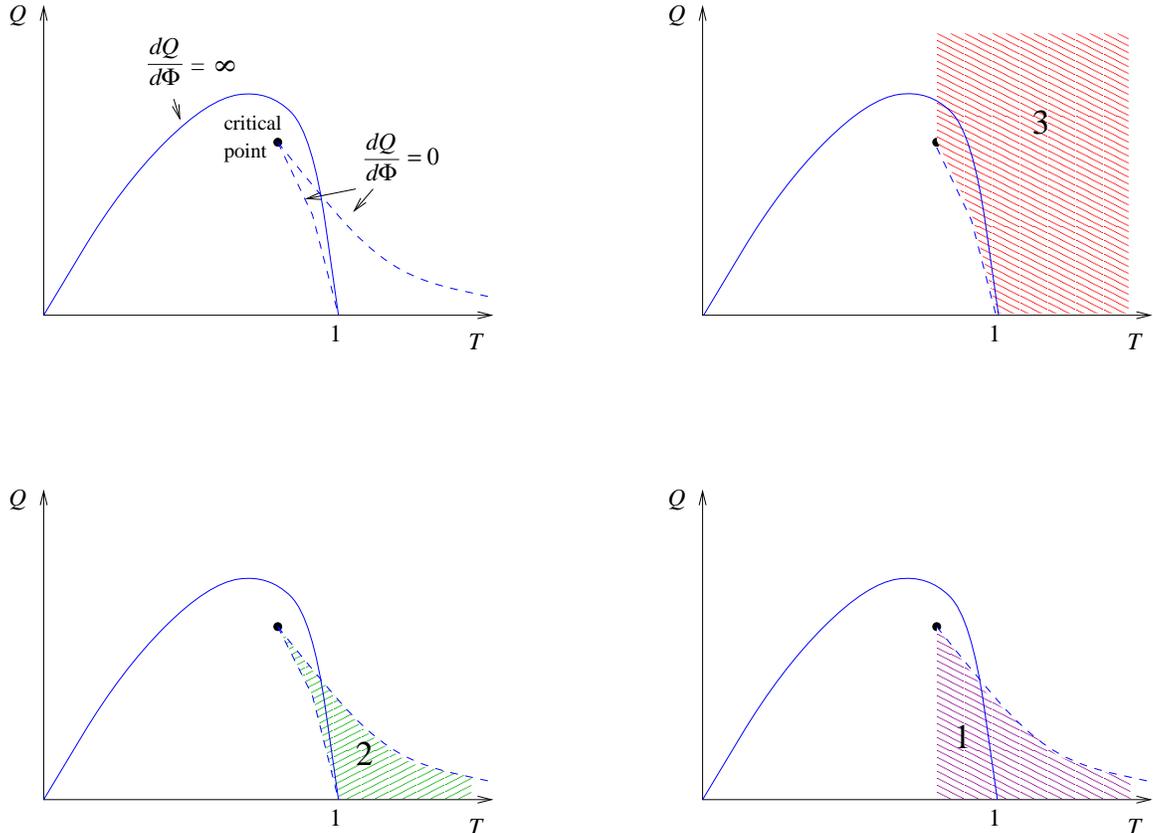,width=6in,angle=270}
\caption{The demarkation of the various branches of black holes in the
$(Q,T)$ plane. Points on branches~1 and~3 which lie inside the solid
curved line are unstable to electric fluctuations. Branch~2 is
electrically stable but thermally unstable everywhere.}
\label{fig:branches}
\end{figure}

\begin{figure}[hb]
\hskip2cm
\psfig{figure=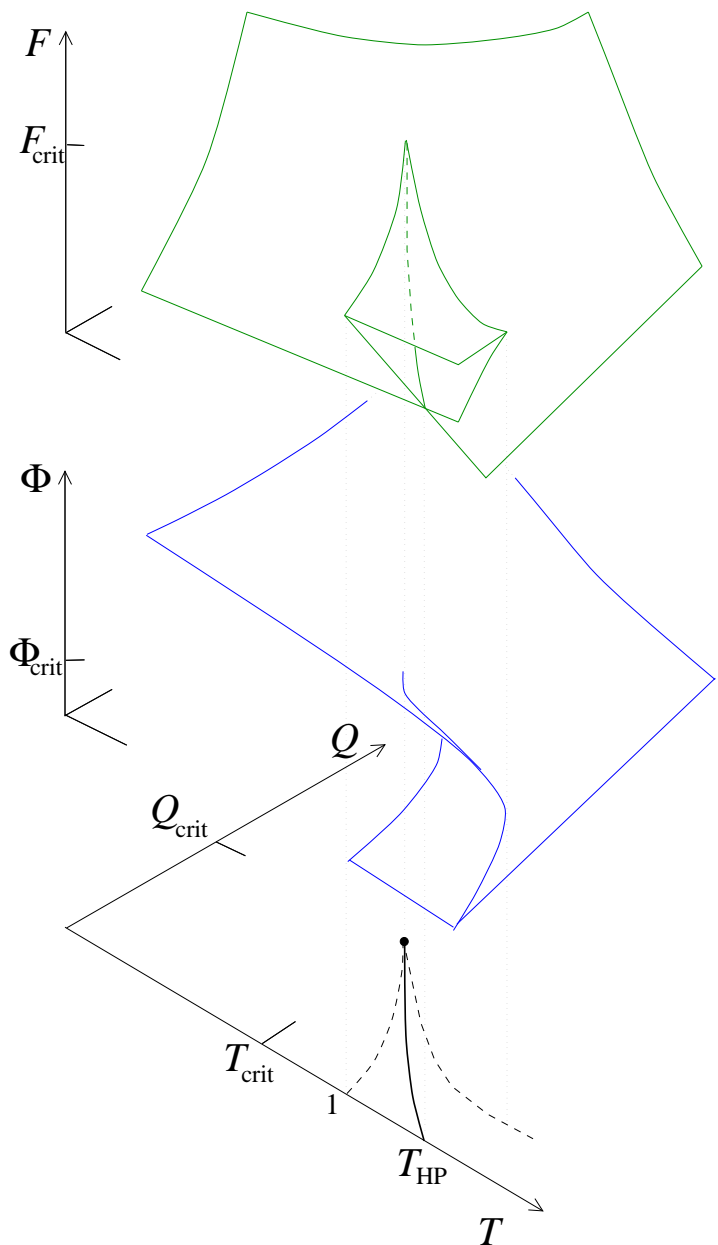,height=6.5in}
\caption{The swallowtail shape (free energy) and cusp shape
(equation of state) for the charged black hole thermodynamic
system. Notice the features which result in the critical line and
point in the $(Q,T)$ plane.}
\label{fig:triple}
\end{figure}

It is particularly interesting to note that the figures in the
previous plot are simply the three sheets of an underlying ``cusp
catastrophe'' shape, as can be seen by assembling them in three
dimensions to reconstruct the equation of state in
figure~\ref{fig:state}. Indeed, it is highly instructive to align the
surface $\Phi(Q,T)$ describing equation of state and the surface
$F[Q,T]$ giving the swallowtail shape of the free energy, in such a
way as to project  some of their important features to
the $(Q,T)$ plane, as done in figure~\ref{fig:triple}. This gives rise
to the critical phase diagram which we will discuss in the next
section.

\begin{figure}[hb]
\hskip3cm
\psfig{figure=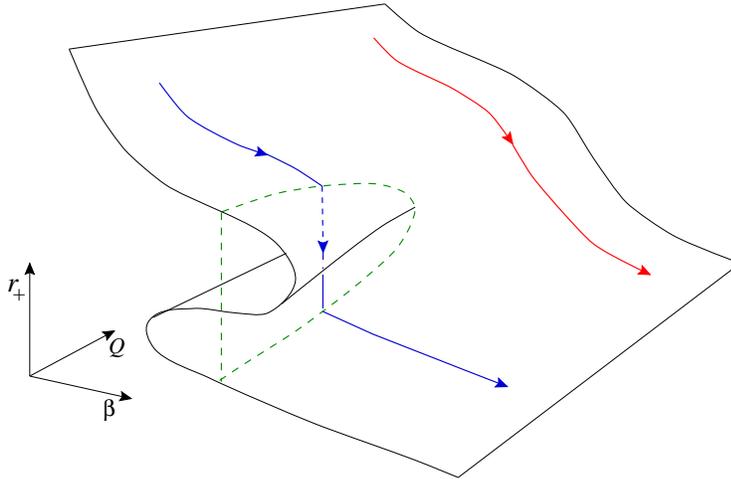,height=2.5in}
\caption{A sketch of the cusp catastrophe  in action (in
$(r_+,Q,\beta)$ space). Two sample
trajectories are shown, one ($Q{<}Q_{\rm crit}$) encountering a phase
transition, the other ($Q{>}Q_{\rm crit}$) does not. The precise
location of the line across which the transition happens is given by
the minimum free energy condition, or equivalently, an appropriately
formulated ``equal area law''.}
\label{fig:cusp}
\end{figure}

As anticipated, the shape formed by the equation of state in the
neighbourhood of the critical point is merely a distortion of the
standard cusp shape, which was encountered in the variables
$(r_+,Q,\beta)$ in our previous
paper~\cite{cejmii}. Figure~\ref{fig:cusp} shows this standard shape
with two sample trajectories in state space. It will be discussed
further in section~\ref{sec:critical}.

As a final comment that in cases where one of the local stability
criteria (\ref{stabcrit}) are violated, we are not always able
to determine the stable ground state. However, the precise nature
of the stability violation is providing information about how the
system will relax to a new stable configuration. For example, one has
$\delta^2E{\propto}\varepsilon_T\,\delta Q^2$ and so $\varepsilon_T{<}0$
indicates that the black hole should relax by reducing its charge,
{\it i.e.,} it will emit charged particles (if possible).

\section{Phase structure}\label{sec:phases}
Figures~\ref{fig:triple} and~\ref{fig:free3d}, together with the
slices displayed in figures~\ref{fig:Tfreesnaps}
and~\ref{fig:Qfreesnaps}, show how the free energy curve determines
the phase structure of the black holes as one moves around on the
state curve in the canonical ensemble, while figures~\ref{fig:gibbs3d}
and the slices displayed in figure~\ref{fig:gibbssnaps} determine the
phase diagram for the grand canonical ensemble.  We performed this
analysis in ref.~\cite{cejmii}, and we recall it here for completeness,
before going on to refine the resulting phase diagram using the
information uncovered in this paper.

The dashed line in the $(Q,T)$ plane shows the boundary of the region
multiply covered by $\Phi$, in the state curve, and correspondingly,
the free energy has three possible values in that region also (see
fig.~\ref{fig:triple}), which constitutes the swallowtail region. The
free energy of branch~2 is always greater than that of either
branches~1 or~3, however, and so there is no transition along the
dashed lines. Along the solid line, the free energies of branches~1
and~3 are equal, and there is a first order phase transition (the
first derivative of the free energy is discontinuous) along this line.
Also note that the one dimensional $Q{=}0$ situation is the familiar
Hawking--Page transition\cite{hawkpage} between AdS and
AdS--Schwarzschild, which happens (in our units) at
$T{=}T_{HP}{=}2/\sqrt{3}{\approx}1.154$, for $n{=}3$.

The solid line is the ``coexistence curve'' of the two phase of
allowed black hole. The line ends in a critical point. Above this
point, there is no transition, and one goes from large to small black
holes continuously (the distinction between branch~1 and branch~3 is
removed). (The reader should compare this to the physics of the
liquid--gas system for an exact analogue in classic thermodynamics.)
As the first derivative (but not the second) of the free energy $F$ is
continuous at the critical point, there is a second order phase
transition there, about which we will have some more to say in
sections~\ref{sec:fluct} and~\ref{sec:critical}.

\begin{figure}[hb]
\hskip1cm
\psfig{figure=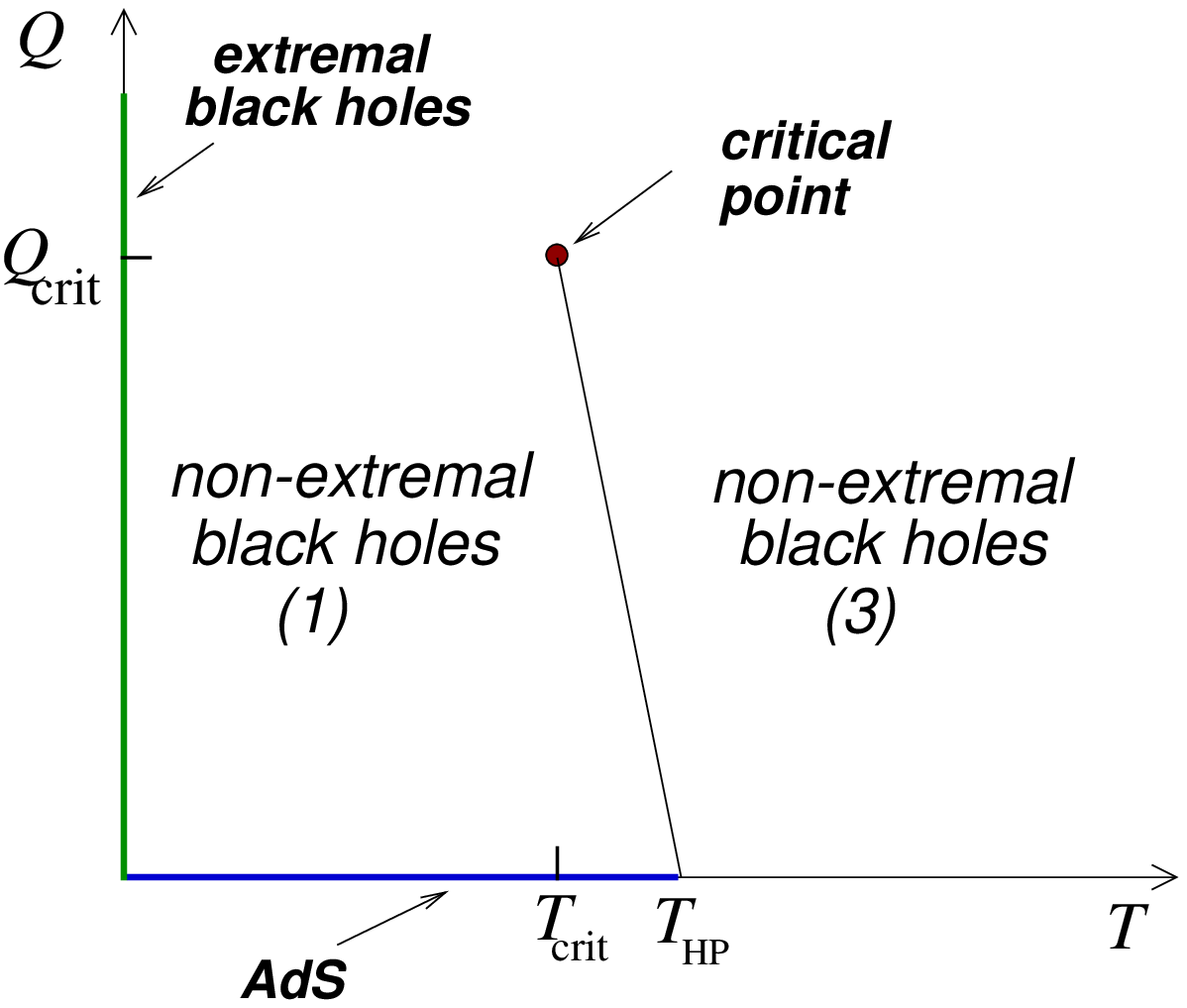,height=2.5in}
\hskip1cm
\psfig{figure=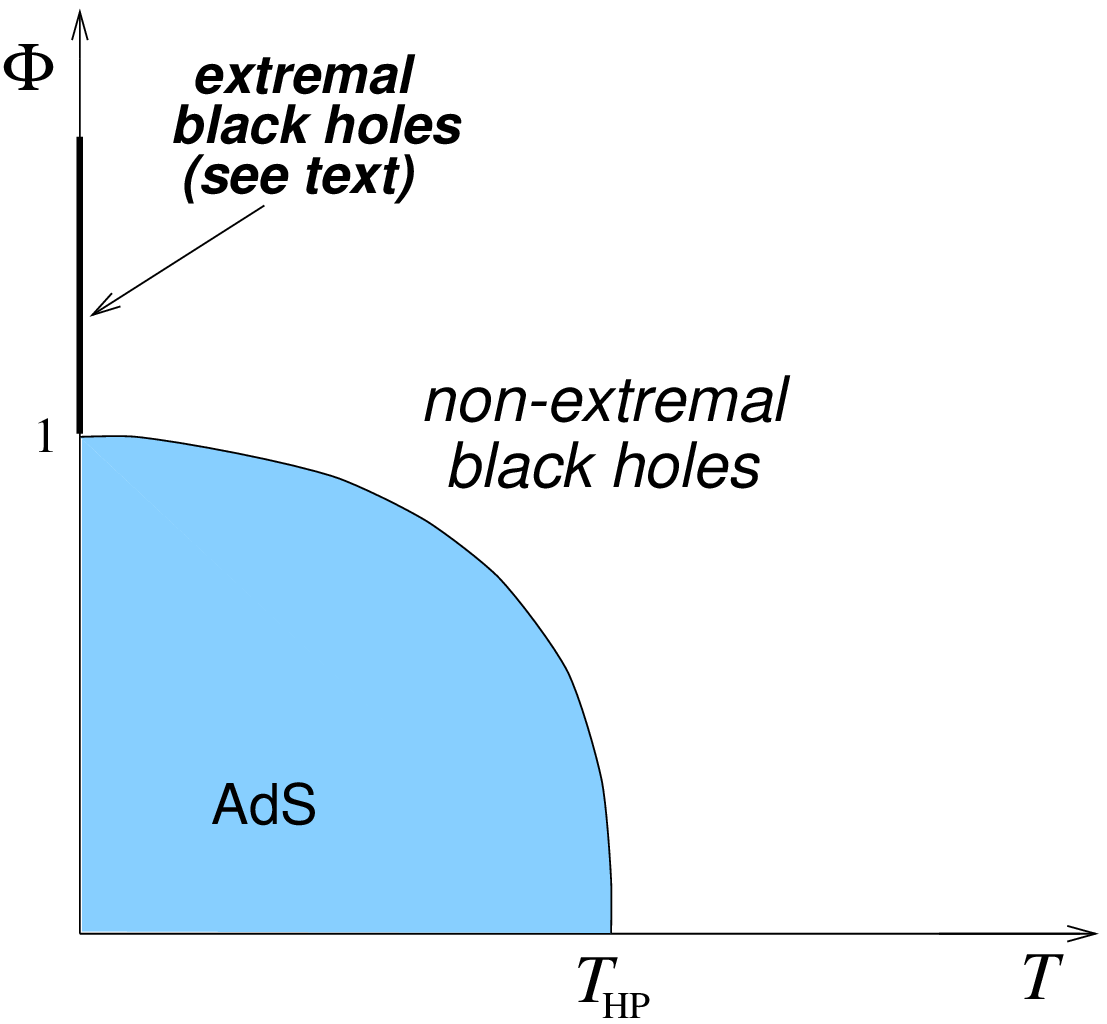,height=2.8in}
\caption{Sketches of the thermally stable phases in the canonical
ensemble, and in the grand canonical ensemble, respectively.}
\label{fig:phases}
\end{figure}

This physics is all summarized in figure~\ref{fig:phases}, where we
have also displayed the phase diagram in the grand canonical ensemble
(the $(\Phi,T)$ plane), which is straightforward to determine. Some of
the details of the shape of these curves will be confirmed by
calculations in section~\ref{sec:coexist}. For most of the rest of the
paper, we will not have much more to say about the phase diagram in
the $(\Phi,T)$ plane, and refer the reader to ref.\cite{cejmii} for
discussions of its features\footnote{Discussed in ref.\cite{cejmii},
for example is the issue of the line of extremal black holes for
$T{=}0$ and $\Phi{>}1$. The calculation of $W[T,\Phi]$ yields a
non--zero result on this line, which is the contribution from the
extremal black holes. We expect that this does not represent the
equilibrium situation, because they will decay due to
``super--radiance'' effects on the approach to zero temperature, as the
charge in them is not fixed in this ensemble. This is an artifact of
the failure of the Euclidean quantum gravity techniques that we have
used to take into account such processes.}.
Note, however, that the boundary in this figure marks the line where
the Gibbs free energy of the black holes equals that of AdS. That is
the boundary does {\it not} denote a curve where one of the local stability
criteria begins to be violated.

Depending upon the situation, there may or may not be the possibility
of electrical fluctuations. This depends very much upon the setting
within which we are considering these black holes. In a theory without
charged particles, the black hole charge would be fixed and electrical
stability need not be considered. 
In general, however, if
there are fundamental charged quanta in the theory, then there is the
possibility of the black holes emitting or absorbing such quanta,
introducing the possibility of electrical fluctuations. Such a
possibility must be considered in (for example) the case when the EMadS
system is considered to be a Kaluza--Klein truncation of some higher
dimensional theory, as discussed in our previous
work\cite{cejmii}. Then, the electrically charged black hole can in
principle emit or absorb electrically charged Kaluza--Klein particles
in order to allow its charge to fluctuate. 

In the particular case of four dimensions, however, there is also the
possibility that we can exchange, by electric--magnetic duality, the
electric charge (and vector potential) that we have been considering
here for a magnetic charge (and vector potential). In this case, we
have instead that the only way for the magnetically charged black
holes to change their charge is to emit or absorb Kaluza--Klein
monopoles, which are not fundamental quanta, as they are very massive,
the further we are below the Kaluza--Klein scale.

In general, when there are allowed electrical fluctuations (by whatever
mechanism is appropriate to the situation in hand) we must also take
into account on the phase diagram, the electrical stability of the
solutions as determined in the previous section. Including those
regions, we obtain the following phase diagram:

\begin{figure}[hb]
\hskip4cm
\psfig{figure=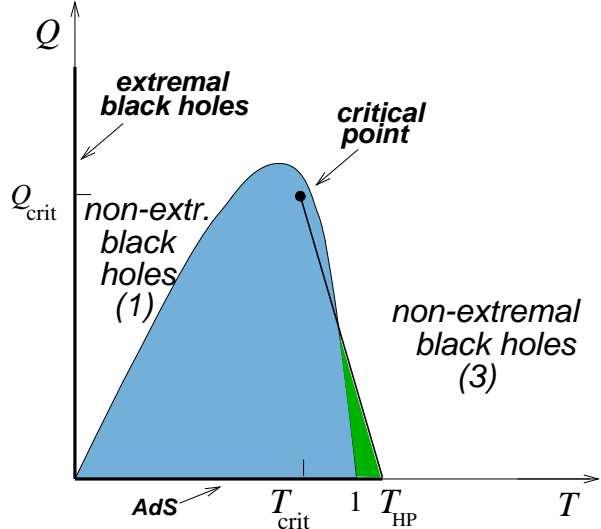,height=3.0in}
\caption{The phase diagram in the canonical ensemble, showing the
disallowed (shaded) regions where the solutions are unstable to
electrical fluctuations. Note that the
critical point and part of the coexistence line lies within the
unstable region.}
\label{fig:critical}
\end{figure}

The question arises as to what the equilibrium system is which resides
in the shaded regions. The electrically unstable black holes cannot
reside there, and so we must search for other possibilities. One
formal possibility is that extremal black holes reside there, because
formally they can exist at any temperature for any charge. However, we
do not find this possibility very attractive. We expect that the
permission that the Euclidean computation appears to give them to
exist at any temperature is an artifact, and that they should
naturally be associated with zero temperature, in which case they can
only occupy the line $T{=}0$ on our phase diagram in the canonical
ensemble, which they do. In any event, one can infer from the
calculations of ref.~\cite{cejmii} that extremal black holes
actually have a higher free energy than the unstable nonextremal
black holes. Another possibility is that the preferred
state is simply anti--de Sitter space (which can also exist at
arbitrary temperature) filled with a charged gas. This is certainly
the case at $Q{=}0$ \cite{edadsii,hawkpage}. However, when the gas carries a
non--negligible charge (and hence mass), its backreaction on the AdS
geometry can not be neglected in determining the free energy.
Another interesting possibility is that of black hole surrounded
by a gas of particles. Again, if the gas component carries a sizable
fraction of the charge and mass, its backreaction on the geometry
would modify the equation of state and may then re--establish
thermodynamic stability. Pursuing either of these possibilities
lies beyond the scope of the present paper, and so
we will leave
settling of this interesting issue to a future date. Hence we must
simply regard
the shaded region as a sort of {\it terra incognita} with regard to
black hole physics. As a final note, we remind the reader that this
is only the region in which we are certain that the black holes do not
minimize the free energy due to its thermodynamic instability.
It may be that the onset of a phase transition to a state of lower
free energy actually occurs outside of the boundary in figure 13,
just as it does for the grand canonical ensemble in figure 12.

\section{Coexistence of phases and the Clapeyron equation}\label{sec:coexist}
Let us study further the coexistence lines which we discovered in the
phase diagrams, both in the canonical ensemble and in the grand
canonical ensemble (see figure~\ref{fig:phases}). We can use
straightforward thermodynamics to determine the shape of the lines.

Let us start with the grand canonical ensemble, with Gibbs potential
$W[\Phi,T]$, with $dW{=}{-}SdT{-}Qd\Phi$.  An equation can be derived
for the line separating two phases A and B in the $(\Phi,T)$ phase
diagram as follows. Along such a coexistence line the phases (for
given $(\Phi,T)$) have the same $W$, and so the slope of the curve
$Q(T)$ is related to the change in entropy and $Q$ by:
\begin{equation}
{d\Phi\over dT} =-{S_{\rm A}-S_{\rm B} \over Q_{\rm A}-Q_{\rm B}}\ .
\end{equation}
In the case at hand, one of the phases is AdS which has zero entropy
and zero charge. So we find that (for all $n$):
\begin{equation}
{d\Phi\over dT} =-{S_{\rm bh} \over Q_{\rm bh}}=-\left({2\pi\over
n-1}\right){1\over cq}\left({q\over c\Phi}\right)^{n-1\over n-2}\ .
\label{ccequation}
\end{equation}
Here, $q(T,\Phi)$ is obtained from the equation of state for the
corresponding branch. Equation~(\ref{ccequation}) is the precise
analogue of the Clapeyron equation. From it, we see that the
slope of the curve is negative.  For the case of $n{=}3$ we can give
explicit expressions. In the rescaled units, we have: 
\begin{equation}
{d\Phi\over dT} =-{Q \over 2\Phi^2}=-{1\over 2}\left({T\over 
\Phi}+\sqrt{{T^2\over \Phi^2}+1 -{1\over \Phi^2}}\right)\ .
\end{equation}
We see from here that the curve intersects the axes orthogonally, and
its convexity, sketched in the figure~\ref{fig:phases}, follows from
the fact that ${d^2\Phi/ dT^2}{<}0$.

Next (assuming the issue of electrical stability can be ignored),
we consider the canonical ensemble, defined by the Helmholtz
thermodynamic potential $F[Q,T]$, with $dF{=}SdT{-}\Phi dQ$. Along any
line of coexistence of two phases, we have:
\begin{equation}
{dQ \over dT} ={S_{\rm A}-S_{\rm B} \over \Phi_{\rm A}-\Phi_{\rm B}}
\end{equation}
The phase diagram is sketched in figure~\ref{fig:phases}. 

The Clapeyron equation can be used to find the slope of the
curve at $Q{=}0$ and $Q{=}Q_{\rm crit}$ for the line separating the
two black holes phases (we show the expressions for all $n$):
\begin{equation}
\left.{dQ \over dT}\right|_{Q_{\rm crit}} =-{\omega_{n-1}\over 4
G}\left({n-1\over n-2}\right) {r_{+(\rm crit)}^{2n-3}\over q_{\rm crit}}\ ,
\end{equation}
\begin{equation}
\left.{dQ \over dT}\right|_{Q{=}0}=-{\omega_{n-1}\over 4 G} r_{+(3)}^{n-1}\ ,
\end{equation}
(where we have used here that $r_{+(1)}^{n-2}\simeq q$ near $q=0$, as
is easy to obtain). On the scale at which we have sketched the
coexistence curve in the previous section, it is essentially a
straight line, and we have drawn it as such in
figure~\ref{fig:phases}.

\section{Fluctuations for Charged AdS Black Holes}\label{sec:fluct}
In section~\ref{sec:stability}, we discussed and computed the
thermodynamic quantities (specific heats and permitivity) which signal
the stability (or not) of a black hole against fluctuations. While
these quantities pertain to the response of the system to macroscopic
thermodynamic processes which may be performed, in Euclidean Quantum
Gravity, where we ordinarily do not have a description of the
microscopic degrees of freedom, we usually cannot relate them directly
to microscopic fluctuations, as we can in ordinary thermal physics.

However, we can go further in this paper. Many of the adS models which
we have here can be embedded into a full theory of quantum gravity
---string and/or M--theory--- and where the holographic duality tells
us precisely that the microscopic description is organized neatly in
terms of a dual (gauge) field theory. 

So we may go and boldly study the fluctuations of the 
thermodynamic quantities in our theory, and we should see earmarks of
the underlying (gauge) theory in our quantities, connecting the
microscopic to the macroscopic.

Here, one uses the entropy to define a probability distribution
on the space of independent thermodynamic quantites~\cite{landau}:
$p(X_i){\propto}\exp(S(X_i))$. With the assumption that fluctuations
are small, we can work with a quadratic expansion of the entropy
in deviations from the equilibrium values. The stability analysis
of section VI establishes that the Hessian of $S$ is negative
semi--definite, and so we have a normalizable Gaussian distribution
within this approximation. One then finds that the fluctuations
are given by
\beq
\langle\delta X_i\,\delta X_k\rangle=
-\left({\partial^2S\over\partial X_i\partial X_k}\right)^{-1}
\label{gauss}
\eeq
where $\delta X_i$ denotes the deviation of $X_i$ from its equilibrium
value, and notation of the left--hand side denotes a matrix inverse.

Implicit above is the assumption that we have a closed system can be divided
into a number of subsystems. In the AdS context, the natural decomposition
is the black hole and the thermodynamic reservoirs\footnote{We are
neglecting the contributions of any gas component around the black hole
in all of our calculations in this paper. Further we should be able
to consider smaller subdivisions with the dual field theory in mind.}. 
In this situation where
the subsystem of interest is really the entire object under study,
the most reasonable approach is to consider fluctuations in only
the extensive variables that are free to vary in the thermodynamic
ensemble \cite{callen}. Hence we denote the
general extensive variables that are free to vary
as $X_i$, and the $F_i$'s are their conjugate
intensive variables defined by $dS{=}F_k\,dX_k$.
Eqn.~(\ref{gauss}) then becomes\cite{callen,landau} 
\beq \langle
\delta X_i\,\delta X_k\rangle=-{\partial X_i\over\partial F_k}
=-{\partial X_k\over\partial F_i}\ .
\label{fluct} \eeq 

 Now, these fluctuations are given practical
meaning when they are compared to for example, their equilibrium
values.  For example, the relative root mean square of the
fluctuations: \beq {\sqrt{\langle\delta X_i^2\rangle}\over X_i} \eeq
which tells us about the sharpness of the distribution in $X_i$. 

Note, that by the formula (\ref{fluct}) the above ratio goes roughly
as the extensive parameters to the ${-}1/2$, and therefore the
distribution is increasingly sharp as the size of the system
increases. 

Now in the present problem of charged black holes,
\beq dS = (1/T)
dE - (\Phi/T) dQ\ .  \eeq
Hence
for the canonical (fixed $Q$) ensemble (our analogue of a fixed volume system),
the only free extensive variable is the energy, and the above formulae
yield 
\begin{equation} \langle \delta E^2 \rangle= -\left({\partial E\over
\partial\beta}\right)_Q =T^2\left({\partial E\over\partial T}\right)_Q=T\,C_Q\ . 
\end{equation}
For the grand canonical (fixed $\Phi$) ensemble (analogue of a fixed
pressure system), the energy and the charge are free to vary, and one
has 
\begin{eqnarray}
 \langle \delta E^2\rangle &=& -\left({\partial
E\over\partial\beta}\right)_{\Phi/T} =T^2\left({\partial
E\over\partial T}\right)_{\Phi/T} \nonumber\\
&=&TC_\Phi+T\Phi\left({\partial E\over\partial \Phi}\right)_T\ . \\
\langle\delta Q^2\rangle&=&T\left({\partial Q\over\partial
\Phi}\right)_T =T\varepsilon_T\ . \\ \langle\delta E\,\delta
Q\rangle&=&T\left({\partial E\over\partial\Phi}\right)_T \nonumber\\
&=&T^2\left({\partial Q\over\partial T}\right)_\Phi
+T\Phi\left({\partial Q\over\partial \Phi}\right)_T=T^2\alpha_\Phi+T\Phi
\varepsilon_T\ .
\end{eqnarray} 
We have recovered the fact that the thermodynamic fluctuations are
controlled by the same generalized compressibilities ---specific heats
permittivity, {\it etc.,}--- that determine the intrinsic stability in
section~\ref{sec:stability}. This follows since both analyses can be
phrased in terms of the Hessian of the entropy.

Above we have presented a general thermodynamic discussion.
Let us now focus on the case $n{=}3$ and present our results in terms
of the dimensionless variables introduced in section ~\ref{sec:intro}.
Note that translating the above thermodynamic formulae to the dimensionless
variables, there are extra factors, giving {\it e.g.,}
\beq \langle\delta  E^2\rangle =
{3G\over2\pi l^2} T^2{\partial 
E\over\partial T}\ . \label{flutter}\eeq
For the fixed charge ensemble: \beq
{\langle\delta E^2\rangle\over E^2}= {3G\over2\pi l^2}{1\over r_+^2}
{\left({\rp^2}+1-{q^2\over\rp^2}\right)^3\over
\left({\rp^2}-1+3{q^2\over\rp^2}\right)
\left({\rp^2}+1+{q^2\over\rp^2}\right)^2}\ . \eeq For the fixed
potential ensemble: \beq {\langle\delta E^2\rangle\over E^2}=
{3G\over2\pi l^2}{1\over r_+^2} {\left({\rp^2}+1-\Phi^2\right)\over
\left({\rp^2}-1+\Phi^2\right)}
\left[{\left({\rp^2}+1-\Phi^2\right)^2+4\Phi^2(1-\Phi^2)\over
\left({\rp^2\over 3}+1-\Phi^2\right)^2}\right]\ , \eeq 
\beq
{\langle\delta Q^2\rangle\over Q^2}= {3G\over4\pi l^2}{1\over\Phi^2
r_+^2}\left[ {\left({\rp^2}+1-\Phi^2\right)
\left({\rp^2}-1+3\Phi^2\right)\over \left({\rp^2}-1+\Phi^2\right)}\right]\ .
\eeq 
\beq {\langle\delta E\delta Q\rangle\over{E\,Q}}= {3G\over\pi
l^2}{1-\Phi^2\over r_+^2}\left[ {\left({\rp^2}+1-\Phi^2\right)\over
\left({\rp^2}-1+\Phi^2\right) \left({\rp^2\over 3}+1-\Phi^2\right)}\right]\ .
\eeq 

Notice that all of these results are proportional to
$G/l^2{\sim}N^{-3/2}$, so for large $N$ the fluctuations are
suppressed.  For $n{=}3$, the dual field theory (supplying our
microscopic description) is the field theory of ref.\cite{exotic},
associated to $N$ coincident M2--branes. The number of degrees of
freedom in this theory grows as $N^{3/2}$ (as seen for example in the
black hole entropy at high temperature). So the squared fluctuations
are controlled by the inverse of the number of degrees of freedom of
the field theory, which is precisely what we expect from standard
kinetic theory connecting the microscopic to the macroscopic! Note
here that we see these unconfined degrees of freedom appearing in our
formulae at arbitrary temperature in this ensemble. This is because
black holes dominate the thermodynamics for all values of the
temperature: the presence of charge effects a deconfinement of the
theory at all temperatures, even in finite volume. (This is to be
contrasted to the case of $Q{=}0$, where AdS dominates the physics for
some $T{<}T_{\rm HP}$, representing the ``confined'' phase.)

To gain more insight into these results, let us 
rewrite eqn.~(\ref{flutter}) for the energy fluctuations as: \beq
{\langle\delta E^2\rangle\over E^2}={3G\over2\pi l^2}{T^3\over r_+^2
\left(T+{Q\over r_+^3}\right)^2}{1\over\left({\partial T\over \partial
r_+} \right)_Q} \eeq 
{}From this form, one can pick out some of the
interesting behaviour. The fluctuations go to zero at zero temperature
as~$T^3$. For large $T$ (and hence large $E$), the fluctuations also
go to zero now as $1/T$ (since for large $r_+$, $2T{\simeq}r_+$). An
interesting factor is $(\partial T/\partial r_+)^{-1}$ which can
change sign for $Q{<}Q_{\rm crit}{=}1/(2\sqrt{3})$.

\begin{figure}[hb]
\hskip4cm
\psfig{figure=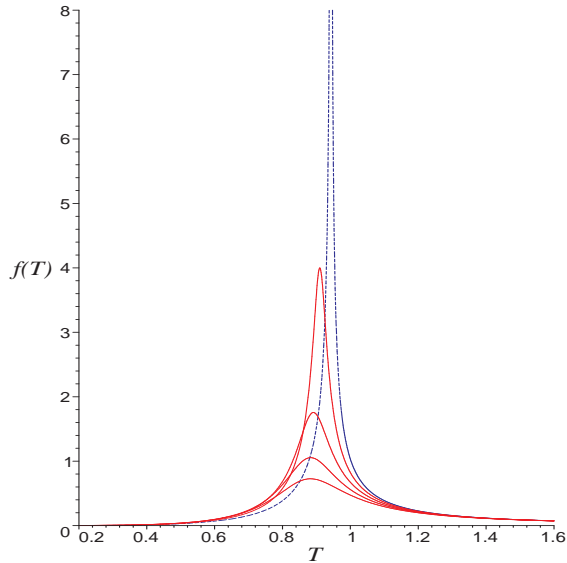,height=3.5in,width=3.5in}
\vskip1cm
\caption{The squared fluctuations $f(T)$ in the energy, relative to the
equilibrium energy, for varying $Q{\geq}Q_{\rm crit}$. The values of
$Q$ plotted here are (bottom up) $Q{=}0.49,0.44,0.39,0.34,Q_{\rm
crit}$. The dotted 
curve shows that the fluctuations diverge at
$Q{=}Q_{\rm crit}$, at the critical temperature $T{=}T_{\rm crit}$.}
\label{fig:fluct}
\end{figure}

So for $Q{>}Q_{\rm crit}$, the fluctuations rise from zero at the
extremal black ($T{=}0$) go through a maximum and then die down for
large temperatures. As $Q$ approaches $Q_{\rm crit}$, the maximum
grows larger and larger, and actually becomes a divergence at
$Q{=}Q_{\rm crit}$!  (We have plotted these squared fluctuations in
figure~\ref{fig:fluct}, where they are denoted $f(T)$.) This is
actually the same divergence as that in $C_Q$ at the critical point,
which was commented on in ref\cite{cejmii}.  Hence one finds from
there that near the critical point, \beq {\langle\delta
E^2\rangle\over E^2}\sim (T-T_{\rm crit})^{-2/3}\ . \eeq

This divergence of the energy fluctuations signals the breakdown
of the Gaussian approximation considered in these calculations.
It is also the classic behaviour of a
second order phase transition point, where correlation lengths, {\it
etc.,} diverge as an order parameter vanishes. Here, the order
parameter can be taken to be a homogenous function of
$r_{+(3)}{-}r_{+(1)}$, the difference between horizon radii of the
branches~3 and~1.

For $Q{<}Q_{\rm crit}$, the fluctuations rise from zero at the
extremal black hole and diverge at the first zero of $\partial
T/\partial r_+$.  Between the two zeros of $\partial T/\partial r_+$,
$\langle\delta E^2\rangle/ E^2$ is negative. This is simply an
indication that we are in the thermally unstable regime, otherwise
known as ``branch~2''.  For $T$ larger than the second zero of
$\partial T/\partial r_+$, the fluctuations are monotonically
decreasing (from infinity at the zero, to zero as
$T\rightarrow\infty$).  As we know from the minimum free energy
condition, we are protected from the unstable regime by making a phase
transition from branch 1 and 3. So in a $\langle\delta E^2\rangle/
E^2$ versus $T$ plot, the fluctuations rise from zero to the phase
transition point and the (discontinuously) jump to the decreasing
curve.

\section{Of Higher dimensions and Other thermodynamic functions 
and ensembles}\label{sec:higher}

In this section, we collect together some results for various
thermodynamic quantities computed for arbitrary $n$, with all of the
factors explicitly included.  The thermodynamic functions are written
in terms of their canonical state variables. We do not use the
physical charge $Q$ instead of $q$, for simplicity of presentation. In
any expression, $Q$ may be restored by recalling that
\begin{equation}
Q={\omega_{n-1}\over 8\pi G}(n-1) c q\quad {\rm and}\quad
c=\sqrt{2(n-2)\over n-1}\ .
\end{equation}
Similarly, we also introduce the parameter $s$ as 
\begin{equation}
S={\omega_{n-1}\over 4G}s\ .
\end{equation}
Notice that it does make sense to write physical quantities in terms
of $q$ and $s$, since they are related to the charge and entropy
densities. This follows from the fact that we may replace
$l^{n-1}\omega_{n-1}$ by the field theory's volume $V_{n-1}$.

The equation of state, following from eqn.~(\ref{betaform}) is:
\begin{equation}
T={(n-2)l^2 (1-c^2\Phi^2)(c\Phi)^{2\over n-2}+ nq^{2\over n-2} \over 
4\pi l^2 (cq\Phi)^{1\over n-2}}
\end{equation}
Equation of state for extremal black holes is:
\begin{equation}
q^{2\over n-2} = {n-2\over n} l^2 (c^2\Phi_e^2-1) (c\Phi_e)^{2\over
n-2},\qquad{\rm for}\quad T\,\, {\rm arbitrary.}
\end{equation}

The Gibbs thermodynamic potential for the grand canonical ensemble
is\cite{cejmii}:
\begin{equation}
W[T,\Phi]={\omega_{n-1}\over 16\pi G l^2}\left[ l^2 {q\over
c\Phi}(1-c^2\Phi^2) -\left({q\over c\Phi}\right)^{n\over n-2}\right]
\end{equation}
where $Q{=}Q(T,\Phi)$ is obtained from the equation of state.  Notice
that $W[T,\Phi]$ vanishes for anti--de Sitter spacetime (which has
$Q{=}0$), and so AdS may be thought of as the reference background for
the calculation of the action, and indeed $W[T,\Phi]$ was computed in
this way in ref.\cite{cejmii}, using the background subtraction
method, which we see (in the present work) gives the same result as
the intrinsic definition by ``counterterm subtraction'' methods.

The Helmholtz free energy which is the Legendre transform of
$W[T,\Phi]$ may be computed with an explicit action calculation, using
the counterterm subtraction method to give an intrinsic
(``backgroundless'') definition. The result is:
\begin{equation}
F[T,Q]={\omega_{n-1}\over 16\pi G l^2}\left[ l^2 {q\over c\Phi}
-\left({q\over c\Phi}\right)^{n\over n-2}+(2n-3)l^2 cq\Phi
\right]\ .
\end{equation}
Again, $\Phi{=}\Phi(T,Q)$ is obtained from the equation of
state. Notice that AdS with with non--zero charge is not a solution of
the equations of motion and so cannot be considered as the ``ground
state'' or reference background for this result. Indeed, this result
cannot be obtained by an action calculation which uses a matching to a
background, precisely for this reason.  The counterterm subtraction
technique is therefore necessary here to supply the honest action
computation for this thermodynamic potential. It is satisfying to note
that $W[T,\Phi]$ and $F[T,Q]$ are Legendre transforms of each other,
$W{=}F{+}Q\Phi$, as they should be.

We can arrive at a variety of other ensembles, with their
corresponding associated potentials, by formal Legendre transforms.
For example, we can consider the enthalpy $H[S,\Phi]$, a function of
entropy and potential (this notation is not to be confused with the
Hamiltonian!). Starting from $W[T,\Phi]$ we can construct
$H{=}W{+}TS$, finding
\begin{equation}
H[S,\Phi]={(n-1)\omega_{n-1}\over 16\pi G l^2}\left[ s^{n\over n-1}
+ l^2 s^{n-2 \over n-1}(1-c^2\Phi^2)\right]\ .
\end{equation}
Notice that this function can {\it not} be obtained by performing a
proper background subtraction in Euclidean gravity, since for any
given solution we cannot find another regular solution with the same
values of the entropy and the potential.  However, the enthalpy
vanishes for AdS, which could therefore be regarded as the ground
state or reference background here.

Another thermodynamic function in terms of its canonical variables is
the (internal) energy $E{=}W{+}TS{+}Q\Phi$,
\begin{equation}
E[S,Q]={(n-1)\omega_{n-1}\over 16\pi G l^2}\left[ s^{n\over n-1}
+ l^2 s^{n-2 \over n-1}+ q^2 l^2s^{-{n-2 \over n-1}}\right]\ ,
\end{equation}
which vanishes as well for AdS. This function would define the
thermodynamic
potential for the microcanonical ensemble, and as for the enthalpy
above, a calculation from
Euclidean gravity should proceed by fixing the entropy of the
state--the black hole area, if we neglect the entropy of the charged
gas in AdS\footnote{See\cite{microc} for work on defining the microcanonical
ensemble in gravity.}.

\section{The Universal Neighbourhood of the Critical Point}
\label{sec:critical}
In the section~\ref{sec:fluct}, we saw that fluctuations diverge as we
approach the critical point in the canonical ensemble. This point
represents a second order phase transition, as can be seen from the
fact that the free energy's first derivative ceases to have a
discontinuity there (see figures~\ref{fig:Tfreesnaps}
and~\ref{fig:Qfreesnaps} for visual confirmation), while the
divergences of the last section signals a discontinuity in the second
derivative.

While much of the detailed discussion of the paper has been for
$n{=}3$, we emphasize here again that the results extend to all~$n{>}2$.
This is most clearly seen from the important features
of the equation of state. Let us examine some of these more closely.

\begin{figure}[hb]
\hskip4cm
\psfig{figure=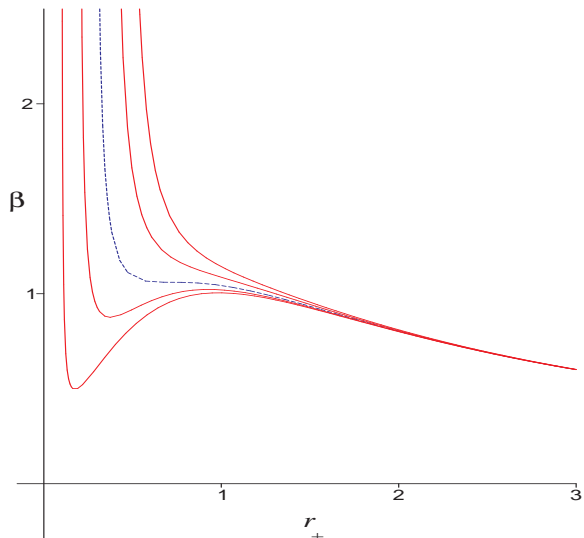,height=2.8in,width=3.0in}
\caption{A family of isocharge curves for the $(\beta,r_{\rm +})$ form
of the equation of state. Note that the middle curve is for the
critical value of the charge, $Q_{\rm crit}$, below which  multiple
branches of $r_{\rm +}$ solutions appear. The neighbourhood of the
critical point is a universal cubic, true for all dimensions.}
\label{fig:isocharges}
\end{figure}

Consider equation~(\ref{betaform}). Originating as the condition for
Euclidean regularity, and hence thermodynamic equilibrium, the
qualitative features of $\beta(r_+)$ for varying $q$ are shown in
figure \ref{fig:isocharges}. These features are the same for all $n$:
There is a critical charge, $q_{\rm crit}$, below which there are
three solutions for $r_+$ for a range of values of $\beta$,
corresponding to the small (branch 1), branch 2, and large (branch 3)
black holes, in the language of ref.\cite{cejmii}, and in this paper.

That this shape persists for arbitrary $n$ can be seen as follows.
First, note that for large $r_+$, $\beta(r_+)$ goes as
${\sim}1/r_+$. Secondly, note that the denominator of the right hand
side of eqn.~(\ref{betaform}) after choosing scalings similar to those
done for $n{=}3$ at the beginning of section~\ref{sec:state},
is\footnote{That is, we absorb a factor of $Gl^{-1}\sqrt{n/(n{-}2)}$
into $Q$ and $l^{-1}\sqrt{n/(n{-}2)}$ into $r_+$, {\it etc.}}
\begin{equation}
r_+^{2n-2}+r_+^{2n-4}-q^2=0\ , 
\end{equation} which has a single positive root,
$r_e$, where $\beta$ diverges. This corresponds to the $T{=}0$
situation, and $r_e$ is the radius of the corresponding extremal black
hole.  Given the above, any turning points for finite $r_+$ must come
in pairs, and the condition $\partial\beta/\partial r_+{=}0$ shows
that there are only two real, positive such solutions, which we call
$r_{+(1)}$ and $r_{+(3)}$, labeling where branch 1 ends and,
respectively, where branch 3 begins.  Branch 2 lies between these
roots.  The equations determining those roots also have an elegant
form (for the same rescaling as before): \bea
r_+^{2n-2}-r_+^{2n-4}+(2n{-}3)q^2=0\ .  \eea

The two roots coalesce at the critical point ({\it i.e.,}
$\partial^2\beta/\partial r_+^2{=}0$ also) where $q{=}q_{\rm
crit}$. The value of the radius of this critical black hole is
$r_{+({\rm crit})}$ and it is at (inverse) temperature $\beta_{\rm
crit}$. For example, in the case $n{=}3$, the quantities $\{q_{\rm
crit},r_{+({\rm crit})},\beta_{\rm crit}\}$ take the values
$\{1/\sqrt{12},1/\sqrt{2},3/(2\sqrt{2})\}$, while for $n{=}4$, they
have the values $\{2/\sqrt{135},\sqrt{(2/3)},5/(4\sqrt{6})\}$.  The
basic point here is that while the critical values themselves vary,
the important structures do not depend upon $n$ in any essential way.

The neighbourhood of the critical point is extremely
interesting. Because of the fact that for all $n$, there at most two
turning points below $q_{\rm crit}$, it is clear that this
neighbourhood can be better written as a cubic, in terms of local
coordinates near the point. To this end, write $\rho{=}r_+{-}r_{+({\rm
crit})}$, ${\hat\beta}{=}\beta{-}\beta_{\rm crit}$, and ${\hat
q}{=}q{-}q_{\rm crit}$, and rewrite the equation of state in these
coordinates.  The neighbourhood of the critical point is found by
taking these coordinates $(\rho,{\hat\beta},{\hat q})$ to be small.

For the example of $n{=}3$, after some algebra, we obtain \be
0=\left(\sqrt{2}-{1\over\beta}\right)\rho^3
+2{\hat\beta\over\beta}\rho^2 +\sqrt{2}{\hat\beta\over\beta}\rho-{\hat
q\over2\sqrt{3}}+{1\over 3}{\hat\beta\over\beta}\ .
\label{thiscubic}
\ee Note that the quadratic and linear terms vanish with at the
approach to the critical temperature ${\hat\beta}{\rightarrow}0$, and
the term which contains $\rho^3$ does not vanish in this way, and so
we neglect higher powers of $\rho$ in favour of this one in order to
study the near--critical behaviour. Here, and in what follows, we will
also neglect terms which are not linear in ${\hat q}$ and $\hat\beta$.
This cubic form~(\ref{thiscubic}) may always be obtained in this limit
for all $n$, because of the observations made in the preceding few
paragraphs.  From this, certain universal behaviour can be easily
deduced, such as the critical exponent characterizing how fast our
``order parameter'', $\rho$, vanishes.  (Recall that $\rho$ represents
the difference in equilibrium radius between the black holes of branch
1 and that of branch 3; it measures the distance from the analogue of
the ``fluid'' phase in liquid--gas language, where the two forms are
indistinguishable.)  Setting ${\hat\beta}{=}0$, we see that the
critical exponent is $1\over3$, since \be
\rho\simeq\left({3\over8}\right)^{1\over6}{\hat
q}^{1\over3}\sim(Q-Q_{\rm crit})^{1\over3}\ .  \ee Performing this
computation for other $n$ will change the numerical prefactor, but not
the exponent, which in this sense deserves to be called universal.

That a cubic equation controls the phase structure can be traced back
a step further. First, notice that the three dimensional plot of the
curve in $(\beta,r_+,Q)$ space is the cusp catastrophe, as drawn in
figure~\ref{fig:cusp}, with some sample state space trajectories.  We
can remove the quadratic term in our cubic polynomial by shifting
$\rho$ by an appropriate amount. Multiplying overall by a normalizing
factor, our cubic may be written as: \begin{eqnarray}
\label{cubic}
0&=&\rho^3+A\rho+B\ , \\ \nonumber
{\rm with}\quad A&\simeq&4\sqrt{2}{\hat\beta}\quad {\rm and}\quad
B\simeq{4\over3}{\hat\beta}-\sqrt{3\over2}{\hat q}\ .
\end{eqnarray} 
Equation~(\ref{cubic}) is actually telling us about the location 
of the turning points
of a {\it quartic} function
\begin{equation}
{\cal V}(\rho)={1\over4}\rho^4+{A\over2}\rho^2+B\rho\ ,
\end{equation}
where we have discarded an arbitrary additive constant.  Treated as a
{\it potential}, (for reasons which will be clear below), it is the
generic form of ${\cal V}(\rho)$ as $A$ and $B$ vary that controls
much of the critical behaviour in the neighbourhood of the critical
point. (As $A$ and $B$ are functions of $Q$ and $T$, this critical
behaviour in $(A,B)$ space translates directly into the earlier
discussed critical behaviour in $(Q,T)$ space.)

The function ${\cal V}(\rho)$ deserves to be treated an effective
potential which organizes the description of much of the local
physics.  In particular, away from the critical point, where $A$ and
$B$ are both non--zero, the potential generically has two minima and
one maximum, the location of which are given by the solutions of our
universal cubic. These locations may be smoothly visualized in the
form of the cusp, sketched in figure~\ref{fig:cusp}. The location of
the minima in $\rho$ are the values $r_{+(1)}$ and $r_{+(3)}$, of the
equilibrium black hole radii of branch 1 and branch 3, while the
location of the maximum is $r_{+(2)}$, the branch 2 black hole
radius. The thermal stability of the branches correlates with the
whether the turning point is a maximum or a minimum of ${\cal
V}(\rho)$, further justifying its treatment as a potential.

The boundary of the region where there are three solutions marks the
situation where one of the minima of the potential ${\cal V}(\rho)$
merges with the maximum and disappears. This boundary is simply given
by the values of $A$ and $B$ where the cubic's discriminant,
$\Delta{=}27B^2{+}4A^3$, vanishes. (This can only happen for $A{<}0$,
therefore telling us that we have the distinct branches {\it below}
$\beta_{\rm crit}$.)  The interior of this region may be translated
into $(Q,T)$ space, where it gives the shaded region in the third
diagram of figure~\ref{fig:branches} where branch 2 resides.

Along the line in the $({\hat q},{\hat\beta})$ plane (or the $(Q,T)$
plane) where $B$ vanishes, given by ${\hat
q}{=}\sqrt{(32/27)}{\hat\beta}$, the two minima of the potential
${\cal V}(\rho)$ are degenerate. This is the point at which there is
phase transition, as the system moves from one minimum of the
potential to the other.

\begin{figure}[hb]
\hskip4.5cm
\psfig{figure=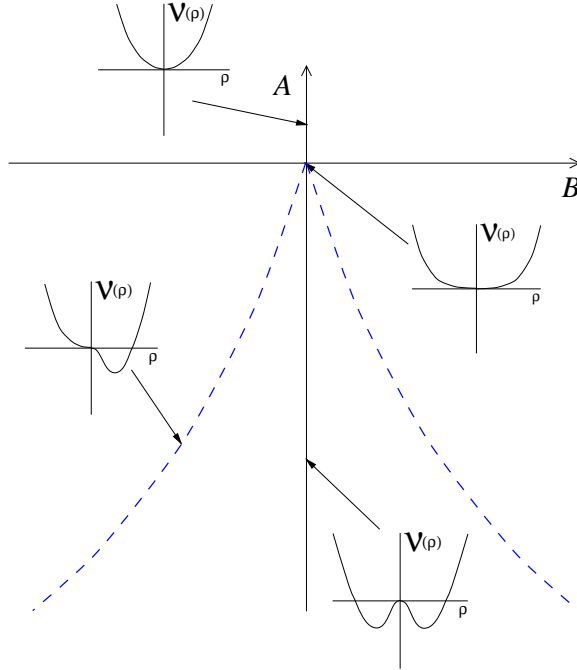,height=3.5in}
\caption{The behaviour of the Landau--Ginzburg $A_3$ potential at various
points in the $(A,B)$ plane. This plane maps to the $(Q,T)$ plane of the
charged black holes system. The line $(A{<}0,B{=}0)$, maps to the critical
coexistence line found in that system.}
\label{fig:LGpot}
\end{figure}

At the critical point, $(A{=}0,B{=}0)$, the maximum and the two minima
merge into a single minimum of the potential. Notice that the well
formed by the potential is very flat there, and so the range of
allowed fluctuations within it is larger at this point than at any
other point in the plane, as they are less confined.  We have seen
this physics before as the divergence of the fluctuations of the
microscopic degrees of freedom at the critical point. The potential
${\cal V}(\rho)$ is an effective potential for the uncharged
microscopic degrees of freedom of the theory in the neighbourhood of
the critical point. (See figure~\ref{fig:LGpot} for a summary of these
critical points of the potential.)

Also, in this language, the meaning of the swallowtail shape for the
thermodynamic potential $F[Q,T]$ is now clear: It is simply the actual
{\it value} of the potential ${\cal V}(\rho;{\hat\beta},{\hat q})$ at
its maxima and minima: the critical line is the place where these two
values at the minima are equal, the place where ${\cal V}$ has
degenerate minima.

This function ${\cal V}(\rho)$ is the $A_3$ Landau--Ginzburg
potential. The effective Landau--Ginzburg theory which we can write
here is an effective theory of the uncharged microscopic degrees of
freedom underlying the thermodynamics. Kinetic terms to complete the
Landau--Ginzburg model would have their origins in the holographically dual
field theory.  One can in principle derive additional potential terms
governing the charged degrees of freedom as well, in order to model
the stability structure uncovered in section~\ref{sec:stability}, but
we will not do that here.

In the language of catastrophe theory\cite{catastrophe}, the term
$\rho^4$ is the basic ``germ'' of the cusp catastrophe, and $A$ and
$B$ are the ``unfolding parameters'' which deform the potential,
giving a line of first order phase transition points along the line
$(B{=}0,A{<}0)$ where its mimima are degenerate.  The ($A$--$D$--$E$)
classification of such potentials is isomorphic to that of certain
geometrical singularities\cite{arnold}. We cannot help but wonder if
this story marks the beginning of a richer tale involving a more
profound underlying geometrical structure into which this physics is
all embedded. As all of the physics of this paper is intimately
connected to the physics of branes, perhaps the possibility of such a
connection should be pursued.


\section*{Acknowledgments}
AC is supported by Pembroke College, Cambridge.  RE is supported by
EPSRC through grant GR/L38158 (UK), and by grant UPV 063.310--EB187/98
(Spain).  Support for CVJ's research was provided by an NSF Career
grant, \# PHY9733173 (UK).  RCM's research was supported by NSERC
(Canada) and Fonds FCAR du Qu\'ebec.  This paper is report \#'s
DAMTP--1999--54, EHU-FT/9907, DTP--99/25, UK/99--5 and
McGill/99--15. RCM would like to thank Martin Grant for interesting
conversations.

\end{document}